# Floquet spin states in OLEDs

S. Jamali[1,⊥)], V. V. Mkhitaryan[1,⊥)], H. Malissa[1], A. Nahlawi[1], H. Popli[1], T. Grünbaum[2], S. Bange[2], S. Milster[2], D. Stoltzfus[3], A. E. Leung[4,†], T. A. Darwish[4], P. L. Burn[3], J. M. Lupton[1,2,*)], and C. Boehme[1,*)]

[1]Department of Physics and Astronomy, University of Utah, Salt Lake City, Utah 84112, USA

[2]Institut für Experimentelle und Angewandte Physik, Universität Regensburg, 93053 Regensburg, Germany

[3]Centre for Organic Photonics & Electronics, School of Chemistry and Molecular Biosciences, The University of Queensland, Brisbane QLD 4072, Australia

[4]National Deuteration Facility, Australian Nuclear Science and Technology Organization (ANSTO), Lucas Heights, New South Wales 2234, Australia

[⊥]equally contributing authors

†Present address: Scientific Activities Division, European Spallation Source ERIC, Lund 224 84 Sweden

*) corresponding authors: john.lupton@ur.de, boehme@physics.utah.edu

Weakly spin-orbit coupled electron and hole spins in organic light-emitting diodes (OLEDs) constitute near-perfect two-level systems to explore the interaction of light and matter in the ultrastrong-drive regime. Under such highly non-perturbative conditions, the frequency at which the spin oscillates between states, the Rabi frequency, becomes comparable to its natural resonance frequency, the Larmor frequency. For such conditions, we develop an intuitive understanding of the emergence of hybrid light-matter states, illustrating how dipole-forbidden multiple-quantum transitions at integer and fractional g-factors arise. A rigorous theoretical treatment of the phenomena comes from a Floquet-style solution to the time-dependent Hamiltonian of the electron-hole spin pair under resonant drive. To probe these phenomena experimentally requires both the development of a magnetic-resonance setup capable of supporting oscillating driving fields comparable in magnitude to the static field defining the Zeeman splitting; and an organic semiconductor which is characterized by minimal inhomogeneous broadening so as to allow the non-linear light-matter interactions to be resolved. The predicted exotic resonance features associated with the Floquet states are indeed found experimentally in measurements of spin-dependent steady-state OLED current under resonant drive, demonstrating that complex hybrid light-matter spin excitations can be formed and probed at room temperature. The spin-Dicke state arising under strong drive is insensitive to power broadening so that the Bloch-Siegert shift of the resonance becomes apparent, implying long coherence times of the dressed spin state with potential applicability for quantum sensing.

*Introduction*

A spin in a magnetic field is a perfect discrete-level quantum system, for which resonant electromagnetic radiation can drive coherent propagation in a well-controlled perturbative fashion following Rabi's theory[1]. This is used for magnetic-resonance applications such as spectroscopy and imaging, where the thermal-equilibrium population of spin states is altered under resonance, inducing an effective magnetization change of nuclear or electronic spins[2]. Such experiments are generally performed under the condition that the Zeeman splitting between spin states induced by a static magnetic field is much larger in energy, or frequency, than the Rabi frequency[1]. This weak-drive limit of resonant pumping is described by perturbation theory. Magnetic resonance can also be probed by observables secondary to magnetization, such as the permutation symmetry of spin pairs of electronic excitations, which is reflected in luminescence or conductivity in atomic, molecular, or solid-state systems[3,4] in optically or electrically detected magnetic resonance (ODMR, EDMR) spectroscopies[5]. Colour centres in crystals, such as atomic vacancies in silicon carbide or diamond, are widely used in ODMR-based quantum metrology and quantum-information processing[6], but are limited in one regard: since dipolar and exchange coupling in the spin pair is strong, substantial level splitting arises at zero external field, posing a lower limit on resonance frequency. Such a limitation does not exist for weakly coupled spin-½ charge-carrier pairs which form, for example, by electron transfer in molecular donor-acceptor complexes, where they account for a range of magnetic-field effects[7].

A particularly versatile way to study weakly coupled spin-½ pairs is offered by OLEDs, which generate light from the recombination of electrically injected electrons and holes that bind in pairs of singlet or triplet permutation symmetry[8]. Since the formation rates of intramolecular excitons from intermolecular electron-hole carrier pairs differ for singlet and triplet spin permutations, and singlet pairs tend to have shorter lifetimes than triplets, an increase of singlet content in the electron-hole pair depletes the reservoir of available carriers and leads directly to a decrease in conductivity[9-11]. Because carrier spins interact with local hyperfine fields, which originate from unresolved hyperfine coupling between charge-carrier spins and the nuclear spins of the ubiquitous protons[12], carrier migration through the active OLED layer gives rise to spin precession and, ultimately, mixing of singlet and triplet carrier-pair configurations. As a result, OLEDs can exhibit magnetoresistance down to sub-microtesla scales at room temperature[5]. A static magnetic field tends to suppress this hyperfine spin mixing partially, an effect which is reversed under magnetic resonance conditions, which give rise to distinct resonances in the magnetoresistance functionality[13].

It is important to stress that such studies are only possible in materials characterized by very weak spin-orbit coupling[14], such as organic semiconductors, and are not generic to electron-hole recombination in LEDs as a whole. In an inorganic LED, for example, spin mixing occurs by spin-orbit coupling in addition to the fact that recombination does usually not occur into tightly bound excitonic species. Light generation in inorganic LEDs, in contrast to OLEDs, is therefore primarily not spin dependent. As such, the subsequent discussion strictly only applies to OLEDs comprising materials of weak spin-orbit coupling, i.e. of low atomic-order number. The appeal of experimenting with spins in OLEDs is that their coherence time, i.e. the transverse spin-relaxation time $T_2$, is only weakly dependent of temperature[15]. This independence is a direct consequence of the lack of spin-orbit coupling, which leaves the carrier dynamics in the local hyperfine fields as the coherence-limiting effect[15].

In principle, magnetic resonances can be resolved down to very small frequencies of a few MHz, limited only by the overall strength of the hyperfine interaction[5,16]. One advantage of EDMR in OLEDs over conventional EPR in radical-pair-based spin-½ systems[17] is that the sample volume can be made almost arbitrarily small. It is therefore possible to achieve high levels of homogeneity both in terms of the static field $B_0$, which defines the Zeeman splitting, and the amplitude of the resonant driving field, $B_1$, while at the same time penetrating the non-perturbative regime of ultrastrong drive where $B_1$ becomes comparable to $B_0$ so that the Rabi frequency approaches the Larmor frequency[18]. Such an ultrastrong-drive regime is of great interest in contemporary condensed-matter physics, although embodiments thereof have proven very challenging to find[19,20] and mostly arise in the form of ultrastrong coupling of two-levels systems in resonant cavities[21-23].

The breakdown of the perturbative regime of OLED EDMR has been identified under drive conditions of $B_1 \approx 0.1 B_0$, where conventional power broadening gives way to a variety of strong-drive effects[24]. Once the driving field strength exceeds the inhomogeneous broadening of the individual spins of the pair induced by the hyperfine fields, the spins become indistinguishable with respect to the radiation and a new set of spin-pair eigenstates is formed[25]. The resonant field locks the spin pairs into the triplet configuration, in analogy to the formation of a subradiant state in Dicke's description of electromagnetic coupling of non-interacting two-level systems[26,27]. The spin-Dicke state manifests itself in EDMR by the appearance of a particular inverted resonance feature[25,28]. Experimentally, it is challenging to access this regime of strong and ultrastrong drive for the simple reason that very high oscillating magnetic field strengths have to be generated in close proximity to the OLED. Using either coils[24] or a monolithic microwire integrated in the OLED structure[18], we were previously able to probe the strong-drive regime of up to $B_1 \approx 0.3 B_0$. We indeed observed an inversion of the resonance signal in the device current along with spectral narrowing due to the spin-Dicke effect. Larger oscillating field strengths were not accessible with these earlier experimental setups. In addition, the magnitude of the $B_0$ field necessary to clearly resolved the resonance was previously limited by the inhomogeneous broadening of the resonance spectrum due to the hyperfine fields arising from the omnipresent protons[16]. Besides limiting spectral resolution, this inhomogeneous broadening also constrains the effective coupling strength of the spin states to the driving field[25]. Qualitatively speaking, the disorder increases quantum-mechanical distinguishability of the spins with respect to the driving field, lowering the overall degree of coherence of the incident radiation with the spin ensemble that can be achieved under resonant drive[25].

To date, there is no theoretical expectation of the nature of spin-dependent transitions of two weakly coupled spin-½ carriers, electron and hole, under ultrastrong resonant drive. Although a Floquet formalism to treat this problem has been put forward previously, this was only pursued in the weak-drive limit[29]. We begin our discussion here by setting out a strategy for computing such transitions in the ultrastrong-drive regime using the periodic time-dependent spin Hamiltonian in the Floquet formalism. The numerically rather detailed calculation can be condensed into a diagrammatic representation of the resulting hybrid light-matter states, the spin states of the OLED dressed by the driving electromagnetic field. The approach gives us a complete computation of the EDMR magnetic resonance spectrum of an OLED under the condition of ultrastrong drive, with $B_1$ exceeding $B_0$. With substantial experimental improvements both in terms of the material used and with regard to the monolithic OLED-microwire structure we succeed in experimentally identifying the main predicted Floquet states of the OLED under ultrastrong drive conditions. These states are manifested as magnetic-dipole-forbidden multiple-quantum transitions and fractional g-factor resonances.

## Results

### Spin transitions in OLEDs under non-perturbative resonant drive

Theoretical treatment of the spin dynamics in a strongly driven electron-hole pair beyond the perturbative regime is extremely challenging and has not been considered in detail previously. We approach this problem using quantum-mechanical Floquet theory[30]. The electron-hole spin-pair Hamiltonian, $H(t)$, is time-dependent due to the presence of a time-dependent (sinusoidal) driving field. Besides the drive, we incorporate in $H(t)$ the electron and hole spin coupling to the external field $B_0$ and the effective local internal hyperfine fields, as well as the isotropic exchange and dipolar interactions between the electron and hole spins. Note that, given the exceedingly weak spin-orbit coupling of the organic semiconductor material used in the experiments discussed in the following, the spin Hamiltonian is perfectly defined without an explicit spin-orbit term, using effective Landé g-factors instead. We direct the interested reader to a recent joint experimental and theoretical examination of the influence of spin-orbit coupling in these materials[14]. While all these interactions are time-independent, the local hyperfine fields and the dipolar interaction are taken to be random and different for different configurations. Further details of the statistical distribution of spin pairs and the numerical approach to account for this are discussed in the Supplementary Information.

The time-periodic character of the driving field allows us to write the Fourier decomposition of the Hamiltonian as $H_{\alpha\beta}(t) = \sum_n H_{\alpha\beta}^{(n)} e^{in\omega t}$. Here, and in the following, the Greek indices $\alpha$ and $\beta$ run over the four spin-pair states, the three triplets $T_+, T_0, T_-$ and the singlet $S$, which are chosen as the spin Hilbert space basis for the subsequent discussion. The Floquet *dressed* states $|\alpha, n\rangle$ with photon number $n = 0, \pm 1, \pm 2, ...$ are defined as the product of the spin Hilbert space basis state $\alpha$ and the Fourier-space basis state $|n\rangle$. The dressed states form an orthonormal basis in an infinite-dimensional *Floquet Hilbert space*. A change of $n$ modifies the dressed state while retaining the same spin-pair component. The time-independent, infinite-dimensional Floquet Hamiltonian $H_F$ is then defined in matrix representation as

$$\langle \alpha, n | H_F | \beta, m \rangle = H_{\alpha\beta}^{(n-m)} + n\omega \delta_{\alpha\beta} \delta_{nm}. \tag{1}$$

The Floquet approach takes advantage of the fact that $H_F$ is time-independent and Hermitian, thus possessing stationary eigenvectors, $|\psi_{\alpha,n}\rangle$, and real eigenvalues, $\varepsilon_{\alpha,n}$. The Floquet Hamiltonian has a periodic structure so that a shift of the indices by the same integer $l$ changes only the diagonal matrix elements

$$\langle \alpha, n+l|H_F|\beta, m+l\rangle = \langle \alpha, n|H_F|\beta, m\rangle + l\omega\delta_{\alpha\beta}\delta_{nm}, \qquad (2)$$

rendering periodicity of the eigenvectors, $\langle \alpha, n+l|\psi_{\alpha,m+l}\rangle = \langle \alpha, n|\psi_{\alpha,m}\rangle$, and eigenvalues, $\varepsilon_{\alpha,n+l} = \varepsilon_{\alpha,n} + l\omega$.

The spin-density matrix of an ensemble of spin pairs, $\rho$, satisfies the stochastic Liouville equation and is used to compute the steady-state singlet content of the spin pair, i.e. the observable responsible for the experimentally measured conductivity of the OLED. Ultimately, this observable can be expressed by the trace of the time-dependent steady-state spin-density matrix, $\mathrm{tr}\tilde{\rho}_0$, where the tilde indicates the solution to the stochastic Liouville equation, which is explicitly time dependent because of the time-periodic Hamiltonian, with the index 0 corresponding to the time-independent component thereof. As described in detail in section S1.2 of the Supplementary Information, from this solution, we can express $\mathrm{tr}\tilde{\rho}_0$ through the following phenomenological parameters: the rate of spin-pair generation, $G$; the spin-pair dissociation rate, $r_d$; the singlet and triplet recombination rates, $r_S$ and $r_T$; and the spin-lattice relaxation time, $T_{\mathrm{sl}}$. For convenience, we introduce a characteristic total decay rate of the pair, $w_d = r_d + r_T + 1/T_{\mathrm{sl}}$, and define the difference in singlet and triplet recombination rates $k_r = r_S - r_T$, which determines the overall magnetic-field response of the conductivity. Using $|\psi_{\alpha,0}\rangle$ as the eigenvectors of the Floquet Hamiltonian and $\Pi_S$ as the projection operator onto the dressed singlet subspace, $\Pi_S = \sum_n |S, n\rangle\langle S, n|$, we arrive at an expression for the time-dependent steady-state spin-density matrix as

$$\mathrm{tr}\tilde{\rho}_0 \simeq \frac{G}{4} \sum_\alpha \frac{1}{w_d + k_r \langle \psi_{\alpha,0} | \Pi_S | \psi_{\alpha,0}\rangle} \ . \tag{3}$$

Equation 3 is derived perturbatively, to leading order in the small difference of singlet and triplet recombination rates $k_r$. Note that $k_r \ll |D|, |J|$, where $D$ and $J$ are the average dipolar and exchange interactions between electron and hole spins within the pairs. Further details of this approximation are discussed in section S1.4 of the Supplementary Information.

The steady-state device current measured in an experiment is a function of the static and oscillating magnetic fields, which give rise to a change $\delta I(B_0, B_1) = I(B_0, B_1) - I(B_0, 0)$ of the spin-dependent OLED current, i.e. the magnetic resonance. This steady-state spin-dependent device current probed in experiments corresponds to the average of $\mathrm{tr}\tilde{\rho}_0$ over all the random distributions of local hyperfine fields and dipolar couplings, i.e. the average over spatial orientations of electronic spin pairs with respect to the applied magnetic field,

$$I = I_0 \langle \mathrm{tr}\tilde{\rho}_0 \rangle, \tag{4}$$

where $I_0$ is a spin-independent factor that is determined by the conductivity of the OLED.

We utilize equations 3 and 4 for the numerical calculation of the EDMR signal $\delta I(B_0, B_1)$. The Floquet Hamiltonian $H_F$ is determined from Eq. 1, for a randomly generated configuration of local hyperfine fields and dipolar coupling. To approximately calculate the infinite-dimensional eigenvectors $\psi_{\alpha,0}$ of $H_F$ entering in Eq. 3, we truncate the infinite-dimensional Floquet Hamiltonian and the corresponding Hilbert space as described in the Supplementary Information. The truncation procedure consists of restricting the photon numbers by some integer valve, $N_0$, or equivalently restricting the indices $m$ and $n$ in Eq. 1 to run between $-N_0$ and $N_0$, where $N_0$ is determined by the requirements on the accuracy of the numerical procedure. The eigenvectors of the truncated Floquet Hamiltonian are used to evaluate $\mathrm{tr}\tilde{\rho}_0$ from Eq. 3. The procedure is repeated multiple times and the field-dependent current is found as the average of the numerical values computed for each random configuration according to Eq. 4.

*Diagrammatic representation of hybrid light-matter states*

In order to build up an intuitive understanding of the resonant transitions of the electron-hole spin pair emerging under strong and ultrastrong resonance drive and to differentiate the Floquet states responsible for the specific transitions, we develop a diagrammatic representation[31] of the dressed states $|\alpha, n\rangle$. For simplicity, we describe the transitions in terms of effective g-factors of resonances. Obviously, this does not relate to the g-factor of the resonant spins – these are all nearly free electrons – but to the ratio between the driving field $B_1$ oscillation frequency and the frequency corresponding to the Zeeman splitting induced by the static magnetic field $B_0$. Figure 1a illustrates the fourfold basis of the Floquet states. A resonant transition involving the creation or annihilation of a photon, shown in Fig. 1b, gives rise to an effective resonance in the overall singlet content of the pair at a *g*-factor $g \approx 2$. This resonance arises from a superposition of transitions depicted by an infinite number of diagrams of increasing order. Specific examples of such diagrams describing the $g \approx 1$ two-photon resonance due to simultaneous annihilation or creation of two photons, shown in Fig. 1c, are given in Fig. 1e. The vortices of the diagrams mark transitions in spin state and occur either by photons, marked as dots, or due to magnetic scattering interactions such as hyperfine or dipolar spin-spin coupling (crosses). The entire singlet content of the spin ensemble in the steady state is computed by approximating the infinite summation over all states (see Supplementary Information). A further group of transitions (Fig. 1d) involves a change of the magnetic quantum number by $\Delta m = \pm 2$, corresponding to a "half-field" resonance at $g \approx 4$. Since the steady-state singlet content is given by a summation over all spin permutations, resonances at fractional $g$-factors will also arise as combinations of the different processes.

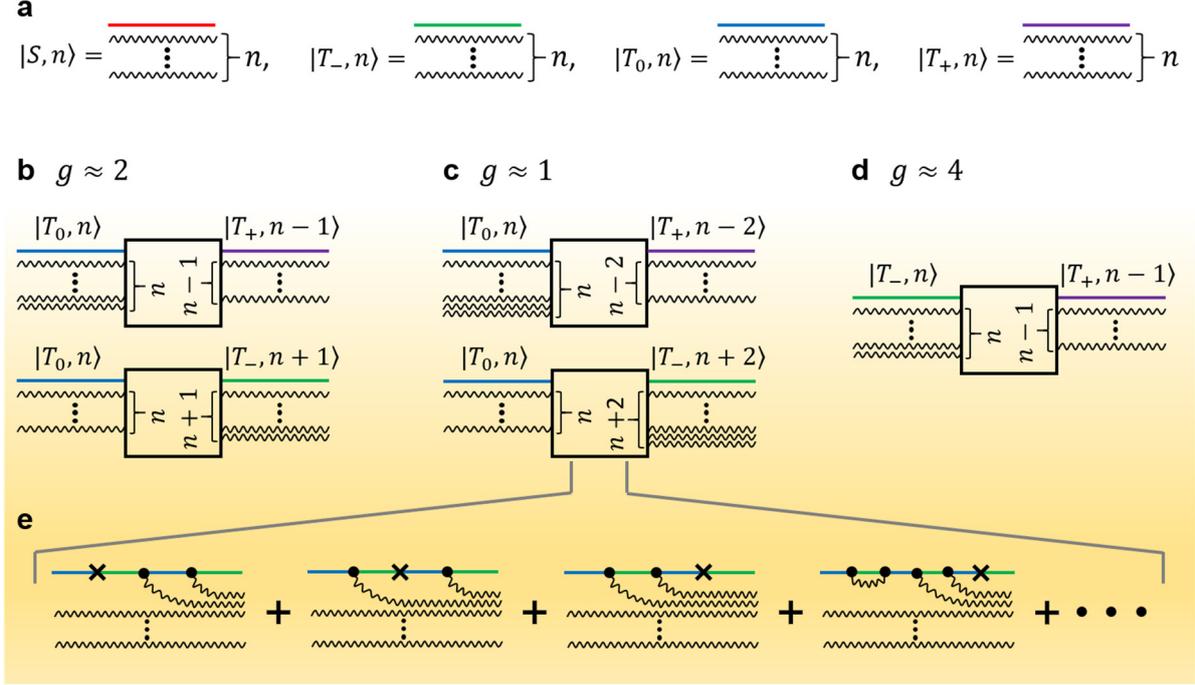

**Fig. 1. Multiple-quantum transitions of spin pairs in the singlet-triplet basis**. *a, The Floquet states describing the conductivity of an OLED under magnetic resonant excitation are defined by the photon number $n$ (illustrated as wavy lines before and after the interaction) and the spin wavefunction (red, green, blue, purple). b, The spin-½ resonance of the pair at an effective g-factor of $g \approx 2$ corresponds to a raising or lowering of $n$. c, d, Resonances also arise at $g \approx 1$ and $g \approx 4$ due to two-photon and half-field transitions, respectively. e, Examples of diagrams of the two-photon transitions, where the vortices indicate the creation or annihilation of a photon (•) or spin scattering not involving a photon (×), e.g. due to hyperfine or dipolar coupling. An infinite number of higher-order loops exists. Magnetoresistance on resonance is calculated by summation over all transitions.*

Even though multi-photon transitions in two-level spin systems have been discussed in the context of electron spin resonance (ESR) spectroscopy[32-34], these are not analogous to two-photon absorption of electric-dipole transitions. Since each photon carries angular momentum $\ell = 1$, a $\Delta\ell = 2$ electric-dipole transition requires a final electronic orbital with different angular momentum. A spin-½ system can only undergo $\Delta\ell = 1$ transitions, however, implying that a magnetic-dipole transition by absorption of two identical photons is impossible. Instead, multi-photon magnetic-dipole transitions arise with a combination of different photons[34] whose magnetic field components have transversal and longitudinal orientation with regard to the magnetic field $B_0$.

*Computation of the EDMR spectrum as a function of driving strength*

In Figure 2a the calculated change of spin-dependent current $\delta I(B_0, B_1)$ as a function of Zeeman splitting $B_0$ and driving field amplitude $B_1$ for an incident radiation frequency of 85 MHz is plotted (see Supplementary Information for numerical details). The width of the fundamental resonance $g \approx 2$ at low driving fields is defined by the expectation value of the local hyperfine field experienced by electron or hole spins, $\Delta B_{hyp}$[25,28], as described in Section S5 of the Supplementary Information. The effective *g*-factors of the resonant species are indicated on the left-hand field ordinate. Six distinct resonances are seen, with the amplitudes and positions of these depending on $B_1$. The six features are assigned the Floquet-state transitions marked in the diagram of spin-pair eigenstates in Fig 2b. Here, states $|2\rangle$ and $|3\rangle$ depend on the particular nature of the intrapair interaction, i.e., dipolar and exchange coupling (see Supplementary Information). Three resonances arise from "full-field" $\Delta m = \pm 1$ transitions, and three from "half-field" $\Delta m = \pm 2$ transitions. Feature (i) is due to the one-photon spin-½ resonance and initially undergoes power broadening, before splitting due to the AC-Zeeman effect and subsequently inverting in amplitude due to the spin-Dicke effect[24]. The $g \approx 1$ feature (ii) results from two-photon transitions, and the $g \approx 2/3$ feature (iii) from three-photon absorption. Resonances also occur between pure triplet states and result from transitions involving either one (iv) ($g \approx 4$), three (v) ($g \approx 4/3$) or five (vi) ($g \approx 4/5$) photons. Features (i)-(iii) exhibit a shift towards lower $B_0$ values with increasing $B_1$, which is a manifestation of the Bloch-Siegert shift (BSS), discussed in more detail below and in the Supplementary Information. The BSS of the $g \approx 2$ resonance results in merging into a zero-field resonance at high $B_1$. The resonances appear self-similar, with the AC-Zeeman and spin-Dicke effects apparent in features (i)-(iii). Crucially, the inversion of $\delta I$ at the onset of the spin-Dicke effect coincides with the formation of a new spin basis of the electromagnetically dressed state[24]. In this hybrid light-matter state, power broadening of the resonant transition is absent since the electromagnetic field is implicit in the state's wavefunction[25], allowing the BSS to be resolved clearly.

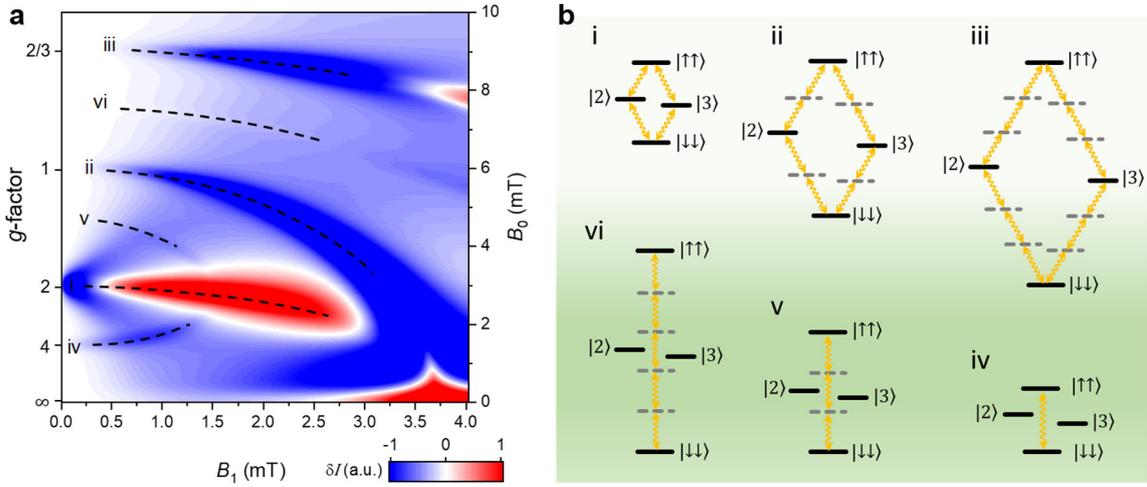

**Fig. 2. Floquet spin states in OLED magnetoresistance.** *a, Calculated change of spin-dependent recombination current as a function of static field $B_0$ and oscillating field $B_1$ for a frequency of 85 MHz. **b**, Term diagrams of integer and fractional g-factor multi-photon transitions.*

*Experimental probing of hybrid light-matter Floquet states in OLED EDMR*

Identifying these Floquet states experimentally in the spin-dependent transitions that control the magnetoresistance of OLEDs poses two challenges. First, the effective hyperfine fields must be sufficiently small, such that the different resonances do not overlap spectrally. Hyperfine coupling broadens the resonant magnetic-dipole transition inhomogeneously, making individual microscopic spins distinguishable in terms of their resonance energy. This disorder determines the threshold field for the onset of spin collectivity[24], when each individual spin becomes indistinguishable with respect to the driving field. Previous studies of the condition of strong magnetic resonant drive of OLEDs employed either conventional hydrogenated organic semiconductors[18,24], or partially deuterated materials with reduced hyperfine coupling strengths[24]. To maximize the resolution of the experiment, we synthesized a perdeuterated conjugated polymer, poly[2-(2-ethylhexyloxy-d$_{17}$)-5-methoxy-$d_3$-1,4-phenylenevinylene-$d_4$] (d-MEH-PPV), with 97% of the protons replaced by deuterons[35]. Second, OLEDs have to be designed with integrated microwires which generate the oscillating field[18]. The smaller the OLED pixel relative to the microwire, which narrows down to a width of 150 μm, the lower the inhomogeneity in both static and oscillating magnetic fields—at the cost of EDMR signal-to-noise ratio. The alternating current passed through the microwire generates heat, requiring careful optimization of electrical and thermal conductivity of the monolithic OLED-microwire device[18].

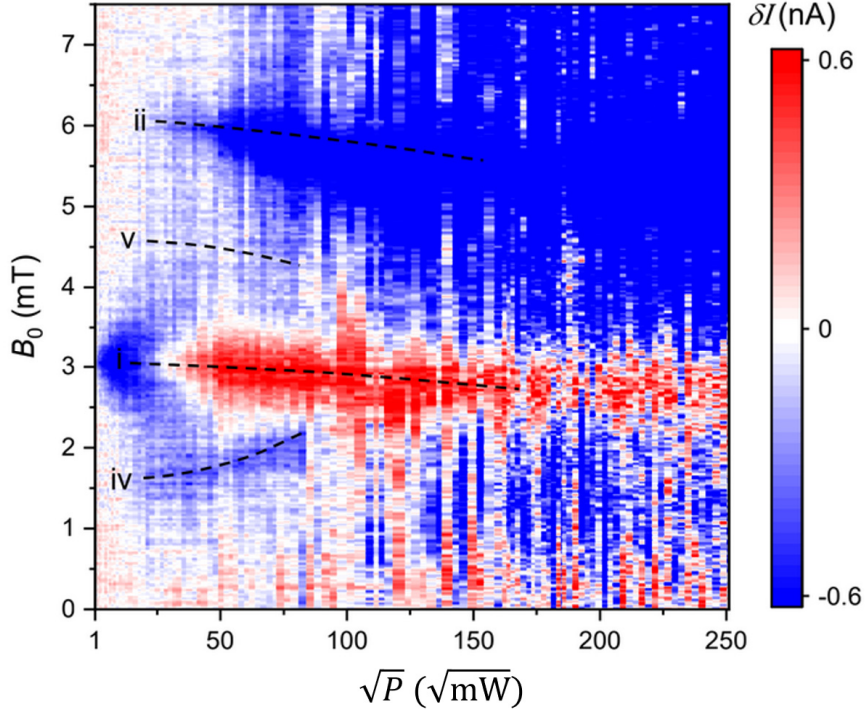

**Fig. 3. Measured change in spin-dependent OLED current δI as a function of driving power.** *The dominant features of Fig. 2 are indicated by dashed lines. The relationship $\sqrt{P} \propto B_1$ is confirmed for $P \leq 100\ mW$ (see Supplementary Information section S4), but, due to heating-induced resistivity changes of the microwire generating the resonant field, the uncertainty on the abscissa scale increases with higher P.*

Figure 3 plots the change $\delta I$ of the steady-state DC forward current $I_0 = 500\ \mu\text{A}$ under 85 MHz RF radiation, as a function of $B_0$ and the square root of the power $P$ of the applied radiation. The dominant Floquet spin state transitions identified in the calculation in Fig. 2 are resolved in the experiment and labelled correspondingly: the power-broadened $g \approx 2$ spin resonance which inverts its sign in the spin-Dicke regime with subsequent BSS (i); the two-photon transition and the corresponding BSS of the $g \approx 1$ resonance (ii); and the half-field resonance at $g \approx 4$ (iv). Some structure is even seen in the spectrum in the range of the resonance at fractional g-factor (v). In principle, $\sqrt{P} \propto B_1$, although heating of the microwire at high powers will change the microwire impedance and therefore increase the uncertainty on the abscissa scale in Fig. 3 (see Supplementary Information). From an analysis of the experimentally observed power broadening given in Fig. S4.1, we estimate that $B_1$ fields of up to 3 mT are reached, consistent with the universal scaling of EDMR amplitude dependence on $B_1$ with hyperfine-field strength as shown in Fig. S7.1. Note that, for experimental reasons, $B_0$ was limited to 7.5 mT. In a separate experiment, we were able to probe the angular dependence of the $g \approx 1$ resonance, offering an additional test of the theory. This two-photon transition is, strictly, dipole forbidden, and only becomes possible because the finite hyperfine-field strength gives rise to oscillating magnetic-field components orthogonal to the plane of polarization of $B_1$ as rationalized in Section S1.6 of the Supplementary Information. In other words, the incident linearly polarized radiation can generate circularly polarized photons at the same frequency, where the magnetic-field vector oscillates in the direction of the axis of quantization[34]. This induced oscillation gives rise to a strong experimental dependence of the two-photon transition intensity between the angle of relative orientation of $B_0$ and $B_1$, $\alpha$, which indeed follows the $\sin 2\alpha$ functionality expected from geometric considerations. We refrain from presenting these data here and will revisit the angular dependence in a future in-depth analysis.

### *Discussion*

We conclude that the Floquet ansatz presented here to solve the time-dependent Hamiltonian of a two-level system in the ultrastrong-drive regime provides both an accurate and surprisingly intuitive representation of the complex spin excitations arising under these conditions. These include the spin-Dicke state, which manifests a strong BSS, along with dressed light-matter states which support dipole-forbidden multiple-quantum transitions. Spin-dependent recombination rates between paramagnetic charge-carrier states in OLEDs offer a remarkably versatile testbed to probe this regime of ultrastrong coupling, visualizing directly the predicted multiple-quantum transitions and fractional g-factor resonances – to our knowledge, better than other known quantum systems at room temperature. The identification of the BSS at room temperature is particularly clear in these solid-state devices compared to any other spectroscopic probe of this phenomenon outside of nuclear magnetic resonance[36-38]. Similarly, direct signatures of hybrid light-matter Floquet states are observed here to influence spin-dependent OLED currents, much like the Floquet states reported in photoelectron spectroscopy of topological materials[39].

The theoretical results of Figure 2 provide motivation to extend the phase space of the experiment to even higher ratios of $B_1/B_0$. While experimentally challenging, this is not entirely unfeasible, and may, for example, be conceivable by using superconducting stripline resonators[9]. It would be particularly exciting to experimentally identify the spin-Dicke state, which is reflected by the inversion and narrowing of the resonance, in the three-photon and two-photon transitions (iii) and (ii) predicted in Figure 2. The latter is especially interesting since the BSS appears to induce a zero-field resonance, i.e. a resonance feature for $B_0 \rightarrow 0$ mT and $B_1 \approx 3.7$ mT.

While we refrain at present from speculating too much on possible technological applications of our study, we do note that a clear feature in theory and experiment is the demonstration that power broadening is suppressed when the spin-Dicke state is formed: the one-photon resonance feature (i) power-broadens linearly with driving field strength $B_1$ when $B_1$ is below the Dicke regime, whereas after inversion of the resonance sign (marked in red in Figures 2 and 3) broadening of the inverted feature with a further increase in power is minimal. This absence of power broadening implies that the hybrid light-matter state, the spin-Dicke state[25], appears to be protected against power broadening precisely because it is a hybrid state. The consequence of this protection should be a dramatically enhanced coherence time, which in turn may turn out to be useful in quantum magnetometry applications. Measuring this increased coherence time in pulsed experiments[40] at very high $B_1$ could potentially confirm this hypothesis. Such experiments would require coherent spin manipulation by RF pulses which are shorter in duration than the RF wave period at the very low Larmor frequencies used here, i.e. subcycle magnetic-resonant coherent control, which is technically feasible with modern pulse synthesizers.

Finally, the work presented here also suggests exciting challenges for materials chemists. The linewidth of the resonances in the calculated spectra of Figure 2 are determined primarily from the finite inhomogeneous broadening arising from the residual hyperfine field strengths of the deuterated conjugated polymer. By careful design of materials, it may be possible to reduce hyperfine coupling even further, for example by extending the π-electron system in polycyclic aromatic hydrocarbons to lower the interaction between electronic and nuclear magnetic moments. With such advances, it will be possible to resolve resonances at even lower $B_0$ fields and therefore reach even higher effective ratios of $B_1$ to $B_0$.

**Methods**

Details of the computational procedure, including the modifications made to the conventional spin-pair model of spin-dependent recombination in an OLED[28,41], are given in the Supplementary Information along with a list of the model parameters used.

In order to establish and detect ultrastrong electron-spin magnetic-resonance drive conditions, we developed a new monolithic OLED device structure with integrated RF microwire. These small circular (57 μm diameter) single-pixel OLEDs allow for bipolar charge carrier injection using electron and hole injector layers on top of an electrically and thermally separated thin-film wire capable of producing RF magnetic fields $B_1$ when connected to an RF source. The wire itself runs across the substrate and is shrunk down to 150 μm width beneath the OLED pixel. Details of the layer sequence necessary to achieve electrical and thermal decoupling are given in Ref. [18]. The active polymer d-MEH-PPV was dissolved in toluene at a concentration of 4.5 g/l at 50-70 °C and deposited by spin-casting at 550 rpm, as described in Ref. [35]. The layer was sandwiched as a 100 nm thin film between a Ca/Al stack for electron injection and a $TiO_2$ (3 nm)/Au (80 nm)/Ti (10 nm) layer covered with PEDOT:PSS (Ossila Al 4083) for hole injection. The structure is comparable to that used in previous studies of the strong-drive regime, where a different π-conjugated polymer with much larger resonance linewidths was used[24]. For the study presented here, the active layer material, the layer stack, and the pixel device geometry were redesigned and optimized in order to allow for larger $B_1/B_0$ ratios to be reached. The monolithic single-pixel OLED device was made following a preparation and deposition protocol as illustrated in Jamali *et al.*[18] with the following changes: (i) the active polymer layer where the electron-hole pair recombination takes place was nominally 100 nm thick, consisting of spin-coated perdeuterated d-MEH-PPV; (ii) the sample template preparation included placing the silicon wafer in a dry thermal oxidation furnace at 1000°C for 78 minutes, producing a 50 nm $SiO_2$ layer on top of the Si wafer for insulation and better adhesion of the subsequent layer (amorphous SiN); and (iii) an entirely different lateral layout of the templates (Figure S2.1a) was used compared to that described in Jamali *et al.*[18], providing larger separation between the electrical contacts of the thin-film wire RF source and the device electrodes in order to avoid cross talk at high RF powers and minimize heating of the OLED during the room temperature measurements. A photograph of the pixel device template is shown in Figure S2.1a with the pixel located at the image centre. An example of the current-voltage characteristics of the device at room temperature is also given in Figure S2.1b.

The OLED was subjected to a static magnetic field $B_0$ and irradiated with RF radiation with in-plane amplitude $B_1$ using an Agilent N5181A frequency generator and an ENI 510L RF power amplifier as illustrated in Figure S2.1a. At the same time, a steady-state electric current was induced with a $V = 2.8$ V bias using a Keithley voltage source. A Stanford Research (SR570) current amplifier was used to detect changes $\Delta I$ of the steady-state forward current of $I_0 = 500$ µA with and without the RF radiation applied. Measurements of $\Delta I(B_0, \sqrt{P})$ as a function of $B_0$ and the square root of the applied microwave power $P$ (which is proportional to $B_1$) were then obtained, as shown in Figure 3. Further details of the measurement procedure and the determination of the $\sqrt{P}$ to $B_1$ conversion factor through power broadening are given in the Supplementary Information in Section S4.

**Data availability**

The raw data that support the plots within this paper and the other findings of this study are available from the corresponding authors upon reasonable request.

**References**

1. Rabi, I. I. Space quantization in a gyrating magnetic field. *Phys. Rev.* **51**, 652-654 (1937).

2. Schweiger, A. & Jeschke, G. *Principles of Pulse Electron Paramagnetic Resonance*, Oxford University Press (2001).

3. Geschwind, S., Collins, R. J. & Schawlow, A. L. Optical detection of paramagnetic resonance in an excited state of $Cr^{3+}$ in $Al_2O_3$. *Phys. Rev. Lett.* **3**, 545-548 (1959).

4. Wrachtrup, J., von Borczyskowski, C., Bernard, J., Orrit, M. & Brown, R. Optical detection of magnetic resonance in a single molecule. *Nature* **363**, 244-245 (1993).

5. Grünbaum, T. *et al*. OLEDs as models for bird magnetoception: detecting electron spin resonance in geomagnetic fields. *Faraday Discuss.* 221, 92 (2020).

6. Doherty, M. W. *et al*. The nitrogen-vacancy colour centre in diamond. *Phys. Rep.* **528**, 1-45 (2013).

7. van Schooten, K. J., Baird, D. L., Limes, M. E., Lupton, J. M. & Boehme C. Probing long-range carrier-pair spin-spin interactions in a conjugated polymer by detuning of electrically detected spin beating. *Nat. Commun.* **6**, 6688 (2015).

8. Pope, M. & Swenberg, C. E. *Electronic Processes in Organic Crystals and Polymers*. (Oxford University Press, New York, 1999).

9. Wang, J., Chepelianskii, A., Gao, F. & Greenham, N. C. Control of exciton spin statistics through spin polarization in organic optoelectronic devices. *Nat. Commun.* **3**, 1191 (2012).

10. Kraffert, F. *et al*. Charge separation in PCPDTBT:PCBM blends from an EPR perspective. *J. Phys. Chem.* C **118**, 28482-28493 (2014).

11. Hoehne, F., Huck, C., Brandt, M. S. & Huebl, H. Real-time electrical detection of coherent spin oscillations. *Phys. Rev. B* **89**, 161305 (2014).

12. Malissa, H. *et al.* Room-temperature coupling between electrical current and nuclear spins in OLEDs. *Science* **345**, 1487-1490 (2014).

13. Kraus, H. *et al.* Visualizing the radical-pair mechanism of molecular magnetic field effects by magnetic resonance induced electrofluorescence to electrophosphorescence interconversion. *Phys. Rev. B* **95**, 241201 (2017).

14. Malissa, H. *et al.* Revealing weak spin-orbit coupling effects on charge carriers in a pi-conjugated polymer. *Phys. Rev. B* **97**, 161201 (2018).

15. Baker, W. J., Keevers, T. L., Lupton, J. M., McCamey, D. R. & Boehme, C. Slow Hopping and Spin Dephasing of Coulombically Bound Polaron Pairs in an Organic Semiconductor at Room Temperature. *Phys. Rev. Lett.* **108**, 267601 (2012).

16. Nguyen, T. D. *et al.* Isotope effect in spin response of pi-conjugated polymer films and devices. *Nat. Mater.* **9**, 345-352 (2010).

17. Maeda, K. *et al.* Chemical compass model of avian magnetoreception. *Nature* **453**, 387-390 (2008).

18. Jamali, S., Joshi, G., Malissa, H., Lupton, J. M. & Boehme, C. Monolithic OLED-microwire devices for ultrastrong magnetic resonant excitation. *Nano Lett.* **17**, 4648-4653 (2017).

19. Günter, G. *et al.* Sub-cycle switch-on of ultrastrong light-matter interaction. *Nature* **458**, 178-181 (2009).

20. Lange, C. *et al.* Extremely Nonperturbative Nonlinearities in GaAs Driven by Atomically Strong Terahertz Fields in Gold Metamaterials. *Phys. Rev. Lett.* **113**, 227401 (2014).

21. Scalari, G. *et al.* Ultrastrong coupling of the cyclotron transition of a 2D electron gas to a THz metamaterial. *Science* **335**, 1323-1326 (2012).

22. Forn-Diaz, P., Lamata, L., Rico, E., Kono, J. & Solano, E. Ultrastrong coupling regimes of light-matter interaction. *Rev. Mod. Phys.* **91**, 025005 (2019).

23. Niemczyk, T. *et al.* Circuit quantum electrodynamics in the ultrastrong-coupling regime. *Nature Phys.* **6**, 772-776 (2010).

24. Waters, D. P. *et al.* The spin-Dicke effect in OLED magnetoresistance. *Nat. Phys.* **11**, 910-914 (2015).

25. Roundy, R. C. & Raikh, M. E. Organic magnetoresistance under resonant ac drive. *Phys. Rev. B* **88**, 125206 (2013).

26. Gross, M. & Haroche, S. Superradiance: An essay on the theory of collective spontaneous emission. *Phys. Rep.* **93**, 301-396 (1982).

27. Dicke, R. H. Coherence in spontaneous radiation processes *Phys. Rev.* **93**, 99-110 (1954).

28. Mkhitaryan, V. V., Danilovic, D., Hippola, C., Raikh, M. E. & Shinar, J. Comparative analysis of magnetic resonance in the polaron pair recombination and the triplet exciton-polaron quenching models. *Phys. Rev. B* **97**, 035402 (2018).

29. Hiscock, H. G., Kattnig, D. R., Manolopoulos, D. E. & Hore, P. J. Floquet theory of radical pairs in radiofrequency magnetic fields. *J. Chem. Phys.* **145**, 124117 (2016).

30. Shirley, J. H. Solution of the Schrödinger equation with a Hamiltonian periodic in time. *Phys. Rev.* **138**, B979-B987 (1965).

31. Cohen-Tannoudji, C., Dupont-Rock, J. & Grynberg, G. *Atom-Photon Interactions: Basic Processes and Applications.* (Wiley, New York, 2004).

32. Clerjaud, B. & Gelineau, A. Observation of electron paramagnetic resonances at multiples of the "classical" resonance magnetic field. *Phys. Rev. Lett.* **48**, 40-43 (1982).

33. Boscaino, R., Gelardi, F. M. & Messina, G. Second-harmonic free-induction decay in a two-level spin system. *Phys. Rev. A* **28**, 495-497 (1983).

34. Gromov, I. & Schweiger, A. Multiphoton resonances in pulse EPR. *J. Magn. Res.* **146**, 110-121 (2000).

35. Stoltzfus, D. M. *et al*. Perdeuteration of poly[2-methoxy-5-(2′-ethylhexyloxy)-1,4-phenylenevinylene] (d-MEH-PPV): control of microscopic charge-carrier spin–spin coupling and of magnetic-field effects in optoelectronic devices. *J. Mater. Chem. C* **8**, 2764-2771 (2020).

36. Forn-Diaz, P. *et al.* Observation of the Bloch-Siegert Shift in a Qubit-Oscillator System in the Ultrastrong Coupling Regime. *Phys. Rev. Lett.* **105**, 237001 (2010).

37. Li, X. W. *et al.* Vacuum Bloch-Siegert shift in Landau polaritons with ultra-high cooperativity. *Nature Phot.* **12**, 324-329 (2018).

38. Sie, E. J. *et al.* Large, valley-exclusive Bloch-Siegert shift in monolayer WS2. *Science* **355**, 1066-1069 (2017).

39. Wang, Y. H., Steinberg, H., Jarillo-Herrero, P. & Gedik, N. Observation of Floquet-Bloch States on the Surface of a Topological Insulator. *Science* **342**, 453-457 (2013).

40. Lupton, J. M., McCamey, D. R. & Boehme, C. Coherent spin manipulation in molecular semiconductors: getting a handle on organic spintronics. *Chem. Phys. Chem.* **11**, 3040 (2010).

41. Boehme, C. & Lips, K. Theory of time-domain measurements of spin-dependent recombination with pulsed electrically detected magnetic resonance. *Phys. Rev. B* **68**, 245101 (2003).

**Acknowledgements**

This work was supported by the US Department of Energy, Office of Basic Energy Sciences, Division of Materials Sciences and Engineering under Award #DE-SC0000909. The synthesis of the perdeuterated MEH-PPV took place at the Australian National Deuteration Facility which is partly funded by NCRIS, an Australian Government initiative. P.L.B. is an Australian Research Council Laureate Fellow and the synthesis work was supported in part by this Fellowship (FL160100067). J.M.L. and S.B. acknowledge funding by the Deutsche

**Author information**

The authors declare that they have no competing financial interests. Correspondence and requests for materials should be addressed to J.M.L. (john.lupton@ur.de) or C.B. (boehme@physics.utah.edu).

**Author contributions**

S.J. designed the high-power microwire OLED structure and performed magnetic resonance spectroscopy, with help from H.M., A.N., and H.P. V.V.M. developed the Floquet simulation code and performed all calculations discussed in the paper. T.G., S.B. and S.M. performed additional supporting experiments at low fields. P.L.B. coordinated the synthesis of the perdeuterated conjugated polymer with support from D.S., A.E.L., and T.A.D. J.M.L. and C.B. conceived and supervised the project, and wrote the paper with input from all authors.

# Floquet spin states in OLEDs

# - Supplementary Information -

S. Jamali[1,⊥)], V. V. Mkhitaryan[1,⊥)], H. Malissa[1], A. Nahlawi[1], H. Popli[1], T. Grünbaum[2], S. Bange[2], S. Milster[2], D. Stoltzfus[3], A. E. Leung[4,†], T. A. Darwish[4], P. L. Burn[3], J. M. Lupton[1,2,*)], and C. Boehme[1,*)]

[1]Department of Physics and Astronomy, University of Utah, Salt Lake City, Utah 84112, USA

[2]Institut für Experimentelle und Angewandte Physik, Universität Regensburg, 93053 Regensburg, Germany

[3]Centre for Organic Photonics & Electronics, School of Chemistry and Molecular Biosciences, The University of Queensland, Brisbane QLD 4072, Australia

[4]National Deuteration Facility, Australian Nuclear Science and Technology Organization (ANSTO), Lucas Heights, New South Wales 2234, Australia

⊥equally contributing authors

†Present address: Scientific Activities Division, European Spallation Source ERIC, Lund 224 84 Sweden

*) corresponding authors: john.lupton@ur.de, boehme@physics.utah.edu

# S1. Calculation of OLED current as a function of $B_0$ and $B_1$

## S1.1 Description of magnetic interactions

The singlet-triplet basis is the natural choice for the formulation of the spin-pair dynamics in the presence of recombination. In an applied static magnetic field, $\boldsymbol{B_0} = B_0\hat{\boldsymbol{z}}$, the electron and hole spin-up $|\uparrow\rangle_e$, $|\uparrow\rangle_h$, and spin-down $|\downarrow\rangle_e$, $|\downarrow\rangle_h$ states occupy the Zeeman levels $\pm\frac{1}{2}\hbar\gamma B_0$, where $\gamma$ is the gyromagnetic ratio of electrons and holes. The singlet-triplet spin-pair states are

$$
\begin{aligned}
|T_+\rangle &= |\uparrow\rangle_e\,|\uparrow\rangle_h, \quad |T_-\rangle = |\downarrow\rangle_e\,|\downarrow\rangle_h, \\
|T_0\rangle &= \frac{1}{\sqrt{2}}(|\uparrow\rangle_e|\downarrow\rangle_h + |\downarrow\rangle_e|\uparrow\rangle_h), \\
|S\rangle &= \frac{1}{\sqrt{2}}(|\uparrow\rangle_e|\downarrow\rangle_h - |\downarrow\rangle_e|\uparrow\rangle_h).
\end{aligned}
\tag{S1}
$$

The Zeeman levels of individual spins can be coupled by a resonant microwave excitation. In the language of triplets and singlets, resonant microwave radiation couples the triplet spin-pair levels. At the same time, nuclear spins create random effective hyperfine magnetic fields at the spin sites inducing interconversion between the singlet and triplet spin-pair levels. Characteristic magnitudes of these hyperfine fields, $b_{\mathrm{hf},e}$ and $b_{\mathrm{hf},h}$, are in general different for electrons and holes, and define two distinct characteristic hyperfine frequencies, $\omega_{\mathrm{hf},\mu} = \gamma b_{\mathrm{hf},\mu}$, with $\mu = e, h$. The statistical distribution of local hyperfine frequencies is presumed to be Gaussian, following the distribution

$$
\mathcal{N}(\omega_{i,\mu}) = \frac{1}{\sqrt{2\pi}\,\omega_{\mathrm{hf},\mu}}\exp\left(-\frac{\omega_{i,\mu}^2}{2\omega_{\mathrm{hf},\mu}^2}\right), \qquad \mu = e, h,
\tag{S2}
$$

where $\omega_{i,e}$, $\omega_{i,h}$, $i = x, y, z$, are the Cartesian components of local hyperfine frequencies $\boldsymbol{\omega_e}$, $\boldsymbol{\omega_h}$, distributed isotropically.

In terms of the electron and hole charge-carrier spin operators, $S_e^i$ and $S_h^i$, $i = x, y, z$, the external and internal static magnetic fields (smf) result in the Hamiltonian

$$H_{\text{smf}} = \omega_0 (S_e^z + S_h^z) + \boldsymbol{\omega_e S_e} + \boldsymbol{\omega_h S_h},$$

where $\omega_0 = \gamma B_0$. The interaction with a microwave (mw) field of amplitude $B_1 = 2\omega_1/\gamma$ and frequency $\omega$ linearly polarized along $\hat{\boldsymbol{x}}$ is given by the Hamiltonian

$$H_{\text{mw}} = 2\omega_1 \cos \omega t \, (S_e^x + S_h^x).$$

To these interactions we add the spin-exchange and dipolar couplings. We assume an isotropic exchange with the simple Hamiltonian

$$H_{\text{ex}} = J \left( \frac{1}{4} - \boldsymbol{S_e S_h} \right).$$

The dipolar coupling requires a more elaborate description. We have

$$H_{\text{dip}} = \sum_{i,k=x,y,z} d_{ik} \, S_e^i S_h^k, \qquad d_{ik} = \frac{\mu_0 \hbar \gamma^2}{4\pi} \langle \frac{r^2 \delta_{ik} - 3r_i r_k}{r^5} \rangle,$$

where $\mu_0$ is the magnetic permeability, $\boldsymbol{r}$ is the vector connecting the electron and hole coordinates, and $\langle ... \rangle$ indicates averaging over the electron and hole wavefunctions. In our calculations, we take the average dipolar interaction energy of 25 neV, corresponding to $D = (\mu_0/4\pi)\hbar\gamma^2 \langle 1/r^3 \rangle \simeq 2\pi \times 6.05$ MHz[1]. We further adopt a simplified picture where the separation of spins within pairs is the same and has a value rendering the average strength $D$. Thus, the dipolar tensor of a spin pair depends on the (randomly oriented) unit vector $\hat{\boldsymbol{r}}$ connecting the two spin sites as $d_{ik} = D(\delta_{ik} - 3\hat{r}_i \hat{r}_k)$. Combining all the above interactions, we arrive at the total spin Hamiltonian

$$H(t) = H_0 + H_{\text{mw}}(t), \quad H_0 = H_{\text{smf}} + H_{\text{ex}} + H_{\text{dip}}, \tag{S3}$$

where $H_0$ is time independent but contains random static interactions. The rationalization of the four eigenstates $|\uparrow\uparrow\rangle = |T_+\rangle$, $|2\rangle = \xi_2|T_0\rangle + \eta_2|S\rangle$, $|3\rangle = \xi_3|T_0\rangle + \eta_3|S\rangle$, and $|\downarrow\downarrow\rangle = |T_-\rangle$ of $H_0$ in Fig. 2b is described in detail elsewhere[2]. In the following, we show that the half-field resonance

features clearly seen in our experiment arise due to the dipolar coupling between the electron and hole spins in a pair. Including the finite exchange interaction in addition allows us to attain a more accurate quantitative agreement between theory and experiment.

## S1.2 Dynamics of a weakly coupled spin-pair ensemble

The spin-density matrix of a spin-pair ensemble, $\rho$, satisfies the stochastic Liouville equation[2,3],

$$\frac{d\rho(t)}{dt} = i[\rho(t), H(t)] + (G/4)\mathbf{1} + \mathcal{R}_{\mathrm{dr}}\{\rho\} + \mathcal{R}_{\mathrm{sl}}\{\rho\}, \tag{S4}$$

where the first term describes the spin dynamics due to the magnetic interactions governed by the time-dependent spin Hamiltonian $H(t)$, $G$ is the spin-pair generation rate, $\mathbf{1}$ the identity operator, $\mathcal{R}_{\mathrm{dr}}$ represents the pair dissociation and recombination, and $\mathcal{R}_{\mathrm{sl}}$ the spin-lattice relaxation processes. The spin-dependent recombination processes are described within the singlet-triplet basis. We assume that the pair dissociation occurs at the equal rate, $r_d$, from all states of a spin pair, whereas the recombination into triplet (T) and singlet (S) excitons occurs with two different constant rates $r_T$ and $r_S$. In terms of the matrix elements we have

$$\mathcal{R}_{\mathrm{dr}}\{\rho\}_{\alpha\beta} = -(r_d + r_T)\rho_{\alpha\beta} - (k_r/2)(\delta_{\alpha S} + \delta_{S\beta})\rho_{\alpha\beta}, \tag{S5}$$

where $\alpha, \beta = T_+, T_0, T_-, S$ enumerate the singlet-triplet spin-pair states, $r_d$ is the carrier-pair dissociation rate, which is assumed to be the same for all spin-pair states, and $k_r = r_S - r_T$ is the difference of singlet and triplet recombination rates, which in our case is presumed to be positive. The spin-lattice relaxation is taken to be of the form

$$\mathcal{R}_{\mathrm{sl}}\{\rho\}_{\alpha\beta} = -(1/T_{\mathrm{sl}})[\rho_{\alpha\beta} - \delta_{\alpha\beta}\mathrm{tr}(\rho/4)]. \tag{S6}$$

The first terms of Eqs. S5 and S6 have the same effect and the corresponding rates are naturally incorporated into the single decay constant,

$$w_d = r_d + r_T + 1/T_{sl}. \tag{S7}$$

The spin-dependent recombination is efficient if the decay rates are smaller than the hyperfine-induced mixing of the spin states, $w_d, k_r \ll \omega_{hf}$. Also note that, besides the spin-lattice relaxation, Eq. S4 implicitly incorporates the $T_1$ and $T_2$ processes originating from the random hyperfine fields. This is believed to be the dominant channel of electronic spin decoherence so one can expect that $T_{sl} \gg T_1, T_2$. In our calculations we assume $T_{sl} \gg 1/\omega_{hf}$, ensuring the compatibility of the two previous conditions.

The charge current measured in a continuous-wave EDMR experiment is proportional to the steady-state free-carrier concentration. The spin-dependent contribution of the spin-pair ensemble to this quantity is determined by the time average

$$\frac{1}{\tau_0} \int_0^{\tau_0} r_d \mathrm{tr}\rho(t)dt, \tag{S8}$$

where $\tau_0$ is the characteristic measurement time. For $\tau_0$ larger than the typical spin-pair decay times, $\tau_0 \gg 1/w_d, 1/k_r$, the integral in Eq. S8 comes from the time domain, where a steady state with the density matrix, $\tilde{\rho}(t)$, is reached. Here, $\tilde{\rho}(t)$ is the steady-state solution of Eq. S4, time-periodic because of the time-periodicity of the Hamiltonian (see Eq. S24). It is then easy to see that, for $\tau_0$ much larger than the time period of the Hamiltonian, the average Eq. S8 converges to $r_d \mathrm{tr}\tilde{\rho}_0$, where $\tilde{\rho}_0$ is the static component of $\tilde{\rho}(t)$. Consequently, the spin-dependent current is found by averaging $\tilde{\rho}_0$ over the random spatial orientations of the dipolar tensor and the random local hyperfine fields (see Eq. 4 in the main text).

### S1.3 Floquet theory approach to the spin-pair Hamiltonian

To find the spin-pair dynamics governed by the Hamiltonian Eq. S3 under a strong microwave excitation of $\omega_1 \sim \omega_0$, we employ the standard quantum-mechanical Floquet theory[4]. We begin with introducing the Floquet *dressed* states $|\alpha, n\rangle$, forming an orthonormal basis in an infinite-dimensional Hilbert space. The Floquet states are enumerated with a pair of indices, where the Greek index runs over the four spin-pair states (i.e. $\alpha = T_+, T_0, T_-, S$ in Eq. S1) and the Latin index is an integer. Using the Fourier decomposition of the time-periodic Hamiltonian Eq. S3,

$$H_{\alpha\beta}(t) = \sum_n H_{\alpha\beta}^{(n)} e^{in\omega t},$$  (S9)

the time-independent, infinite-dimensional Floquet Hamiltonian $H_F$ is defined in the matrix representation by Eq. 1 of the main text. The operator $H_F$ is Hermitian and possesses real eigenvalues and orthonormal eigenvectors,

$$H_F |\psi_{\alpha,n}\rangle = \varepsilon_{\alpha,n} |\psi_{\alpha,n}\rangle.$$  (S10)

From the definition Eq. 1 it follows that the Floquet Hamiltonian has a periodic structure. In particular, with a shift of the integer indices by the same integer $k$, only the diagonal matrix elements are changed (see Eq. 2 in the main text). The secular equation $\det(H_F - \varepsilon\mathbf{1})$, from which $\varepsilon_{\alpha,n}$ are found, is unchanged if $\varepsilon$ is replaced by $\varepsilon + k\omega$. Thus, if $\varepsilon$ is an eigenvalue, $\varepsilon + k\omega$ is also an eigenvalue, meaning that one can label the eigenvalues as $\varepsilon_{\alpha,n} = \varepsilon_\alpha + n\omega$, where $\varepsilon_\alpha = \varepsilon_{\alpha,0}$ is chosen between 0 and $\omega$. Furthermore, the components of the eigenvectors obey the periodicity relation

$$\langle \alpha, n+k | \psi_{\alpha,m+k} \rangle = \langle \alpha, n | \psi_{\alpha,m} \rangle.$$  (S11)

## S1.4 Formal solution of the stochastic Liouville equation

In the subsequent analysis of the stochastic Liouville equation S4 we exploit the non-Hermitian Hamiltonian

$$\mathcal{H}(t) = H(t) - i(w_d/2)\mathbf{1} - i(k_r/2)\mathrm{P}_S,$$  (S12)

where $\mathrm{P}_S = |S\rangle\langle S|$ is the projection onto the singlet state. The complex Hamiltonian $\mathcal{H}(t)$ incorporates the first and the third operators of Eq. S4, as well as the first term of the spin-lattice relaxation, Eq. S6. In terms of the non-Hermitian Hamiltonian, the Liouville equation is rewritten as

$$\frac{d\rho(t)}{dt} = i(\rho\mathcal{H}(t) - \mathcal{H}^\dagger(t)\rho) + \frac{1}{4}\left(G + T_{\mathrm{sl}}^{-1}\mathrm{tr}\rho\right)\mathbf{1}.$$  (S13)

From simple population gain and loss arguments one can see that, under steady-state conditions, the total spin-pair population is restricted by $\mathrm{tr}\rho \lesssim G/4w_d$ (recall that $r_S > r_T$). Thus, for long spin-lattice relaxation times $4T_{\mathrm{sl}}w_d \gg 1$ the $\mathrm{tr}\rho$ term in Eqs. S6 and S13 can be regarded as a small correction to the source term. We take advantage of this fact and neglect the term in Eq. S13. Below, we check numerically that in the parametric domain of interest the effect of this term is insignificant. Note that the effect of the spin-lattice relaxation is still preserved through the first term of Eq. S6. After dropping the last term of Eq. S13 we write its formal solution as

$$\rho(t) = U(t, t_0)\rho(t_0)U^\dagger(t, t_0) + \frac{G}{4}\int_{t_0}^{t} dt' V(t, t'),$$  (S14)

where $V(t_1, t_2) = U(t_1, t_2)U^\dagger(t_1, t_2)$ is introduced, and the time-evolution operator is defined in terms of the time-ordered exponential

$$U(t_1, t_2) = T \exp\left[-i\int_{t_2}^{t_1} dt' \mathcal{H}(t')\right].$$  (S15)

Equations S11 and S12 are treated within the quantum-mechanical Floquet theory[4]. The dressed states $|\alpha, n\rangle$ form an orthonormal basis in the infinite-dimensional Hilbert space of the Hermitian Floquet Hamiltonian $H_F$, Eq. 1. The Floquet counterpart of the non-Hermitian Hamiltonian $\mathcal{H}(t)$, defined by

$$\langle\alpha, n|\mathcal{H}_F|\beta, m\rangle = \langle\alpha, n|H_F|\beta, m\rangle - i(w_d/2)\delta_{\alpha\beta}\delta_{nm} - i(k_r/2)\delta_{\alpha S}\delta_{S\beta}\delta_{nm},$$  (S16)

acts in the same Hilbert space. Matrix elements of the time-evolution operator can be written as

$$U_{\alpha\beta}(t_1, t_2) = \sum_n \langle \alpha, n | e^{-i\mathcal{H}_F(t_1 - t_2)} | \beta, 0 \rangle e^{i\omega nt} . \tag{S17}$$

Because of the non-Hermitian character of $\mathcal{H}_F$, its right- and left-hand eigenvectors are not the Hermitian conjugates of each other and must be considered independently. For the right-hand eigenvectors we use the ket vectors

$$\mathcal{H}_F | \lambda_{\alpha,n} \rangle = \chi_{\alpha,n} | \lambda_{\alpha,n} \rangle, \tag{S18}$$

whereas the left-hand eigenvectors are denoted by bra vectors

$$\langle \lambda_{\alpha,n} | \mathcal{H}_F = \chi_{\alpha,n} \langle \lambda_{\alpha,n} |. \tag{S19}$$

By normalizing the eigenvectors one can form an orthonormal set with respect to the non-Hermitian scalar product, $\langle \lambda_{\alpha,n} | \lambda_{\beta,m} \rangle = \delta_{\alpha\beta} \delta_{nm}$. Taking the Hermitian conjugate of Eqs. S18 and S19 we confirm that the right-hand eigenvectors of $\mathcal{H}_F^\dagger$ are the Hermitian conjugates of the left-hand eigenvectors of $\mathcal{H}_F$, and *vice versa*. We have

$$\mathcal{H}_F^\dagger | \bar{\lambda}_{\alpha,n} \rangle = \chi_{\alpha,n}^* | \bar{\lambda}_{\alpha,n} \rangle, \qquad | \bar{\lambda}_{\alpha,n} \rangle = \langle \lambda_{\alpha,n} |^\dagger, \tag{S20}$$

where the asterisk indicates the complex conjugate, and

$$\langle \bar{\lambda}_{\alpha,n} | \mathcal{H}_F^\dagger = \chi_{\alpha,n}^* \langle \bar{\lambda}_{\alpha,n} |, \qquad \langle \bar{\lambda}_{\alpha,n} | = | \lambda_{\alpha,n} \rangle^\dagger. \tag{S21}$$

Note that the above eigenvalues have imaginary parts of a definite sign, since

$$\Im m(\chi_{\alpha,n}) < 0. \tag{S22}$$

The periodicity property Eq. S11 translates into

$$\langle \alpha, n+k | \lambda_{\alpha,m+k} \rangle = \langle \alpha, n | \lambda_{\alpha,m} \rangle, \qquad \langle \lambda_{\alpha,m+k} | \alpha, n+k \rangle = \langle \lambda_{\alpha,m} | \alpha, n \rangle, \qquad \text{(S23)}$$

with similar relations holding for $|\bar{\lambda}_{\alpha,n}\rangle$ and $\langle \bar{\lambda}_{\alpha,n}|$. Using the periodicity properties, the integrand operator of Eq. S14 is given by

$$V_{\alpha\beta}(t_1, t_2) = \sum_n \left\langle \alpha, n \left| e^{-i\mathcal{H}_F(t_1-t_2)} e^{i\mathcal{H}_F^\dagger(t_1-t_2)} \right| \beta, 0 \right\rangle e^{i\omega n t}. \qquad \text{(S24)}$$

Furthermore, the normalized eigenvectors ensure the partitions of unity, $\sum_{\alpha,n} |\lambda_{\alpha,n}\rangle\langle \lambda_{\alpha,n}| = \mathbf{1}$, and $\sum_{\alpha,n} |\bar{\lambda}_{\alpha,n}\rangle\langle \bar{\lambda}_{\alpha,n}| = \mathbf{1}$. Using these properties together with Eqs. S18-S21, we rewrite Eq. S24 as

$$V_{\alpha\beta}(t_1, t_2) = \sum_{n,\nu,k,\mu,p} \langle \alpha, n | \lambda_{\nu,k} \rangle \langle \lambda_{\nu,k} | \bar{\lambda}_{\mu,p} \rangle \langle \bar{\lambda}_{\mu,p} | \beta, 0 \rangle e^{-i(\chi_{\nu,k}-\chi_{\mu,p}^*)(t_1-t_2)} e^{i\omega n t_1}. \qquad \text{(S25)}$$

The integral in the solution Eq. S14 can now be taken using the above relation. We find

$$\int_{t_0}^t dt' V_{\alpha\beta}(t, t') = \sum_{n,\nu,k,\mu,p} \langle \alpha, n | \lambda_{\nu,k} \rangle \langle \lambda_{\nu,k} | \bar{\lambda}_{\mu,p} \rangle \langle \bar{\lambda}_{\mu,p} | \beta, 0 \rangle \frac{1 - e^{-i(\chi_{\nu,k}-\chi_{\mu,p}^*)(t-t_0)}}{i(\chi_{\nu,k}-\chi_{\mu,p}^*)} e^{i\omega n t}. \qquad \text{(S26)}$$

We note that the first term in Eq. S14 represents a transient contribution to $\rho(t)$, vanishing after a time interval $(t-t_0) \gtrsim \max(1/w_d, 1/k_r)$, so that it does not contribute to the steady-state density matrix $\tilde{\rho}(t)$ which we are looking for. Moreover, due to the negative imaginary parts of the eigenvalues, Eq. S22, the term $\exp\left[-i(\chi_{\nu,k}-\chi_{\mu,p}^*)(t-t_0)\right]$ in Eq. S26 decays exponentially for time intervals of the same order and thus not contribute to $\tilde{\rho}(t)$ either. Thus, we arrive at the steady-state density matrix

$$\tilde{\rho}_{\alpha\beta}(t) = \frac{G}{4} \sum_{n,\nu,k,\mu,p} \frac{\langle \alpha, n | \lambda_{\nu,k} \rangle \langle \lambda_{\nu,k} | \bar{\lambda}_{\mu,p} \rangle \langle \bar{\lambda}_{\mu,p} | \beta, 0 \rangle}{i(\chi_{\nu,k}-\chi_{\mu,p}^*)} e^{i\omega n t}. \qquad \text{(S27)}$$

Equation S27 is the Fourier decomposition of the steady-state density matrix $\tilde{\rho}(t)$, which is periodic with the period of the drive field. In Eq. S8, which defines the contribution to the spin-dependent current $I$, the oscillatory components of $\tilde{\rho}(t)$ average to zero as the integration time is large, $\tau_0 \gg 2\pi/\omega$. Thus, the only contribution to $I$ comes from the time-independent component,

$$\tilde{\rho}_{0,\alpha\beta} = \frac{G}{4} \sum_{\nu,k,\mu,p} \frac{\langle \alpha,0|\lambda_{\nu,k}\rangle\langle\lambda_{\nu,k}|\bar{\lambda}_{\mu,p}\rangle\langle\bar{\lambda}_{\mu,p}|\beta,0\rangle}{i(\chi_{\nu,k} - \chi_{\mu,p}^*)}.$$  (S28)

Using this relation with the periodicity properties, Eq. S23, we arrive at

$$\operatorname{tr}\tilde{\rho}_0 = \frac{G}{4} \sum_{\alpha,\nu,k} \frac{\langle \lambda_{\nu,k}|\bar{\lambda}_{\alpha,0}\rangle\langle\bar{\lambda}_{\alpha,0}|\lambda_{\nu,k}\rangle}{i(\chi_{\nu,k} - \chi_{\alpha,0}^*)}.$$  (S29)

We find the eigenvalues of $\mathcal{H}_F$ and $\mathcal{H}_F^\dagger$ perturbatively with respect to small $k_r$, by splitting these non-Hermitian operators into the unperturbed parts equivalent to the Hermitian Floquet Hamiltonian $H_F$ and the perturbation parts $\pm i(k_r/2)\Pi_S$:

$$\mathcal{H}_F = H_F - i(k_r/2)\Pi_S - i(w_d/2)\mathbf{1}, \qquad \mathcal{H}_F^\dagger = H_F + i(k_r/2)\Pi_S + i(w_d/2)\mathbf{1}$$  (S30)

where $\Pi_S = \sum_n |S,n\rangle\langle S,n|$ is the projection operator onto the dressed singlet subspace. Using quantum-mechanical perturbation theory, we find

$$\chi_{\alpha,n} \approx \varepsilon_{\alpha,n} - i(k_r/2)\langle\psi_{\alpha,n}|\Pi_S|\psi_{\alpha,n}\rangle - i(w_d/2),$$  (S31)

where $|\psi_{\alpha,n}\rangle$ and $\varepsilon_{\alpha,n}$ are the eigenvectors and eigenvalues of $H_F$, Eq. S10. We further note that $\langle\lambda_{\nu,k}|\bar{\lambda}_{\mu,p}\rangle = \delta_{\nu\mu}\delta_{kp} + \mathcal{O}(k_r)$, so the sum in Eq. S29 is dominated by terms with $\nu = \alpha$ and $k = 0$. To the leading order, the numerators of these terms can be replaced by 1. We keep only these terms to arrive at the expression Eq. 3 in the main text.

The approximation leading from the exact relation Eq. S29 to the perturbative result Eq. 3 is valid when the level spacing of any two stationary states $|\psi_{\alpha,n}\rangle$ and $|\psi_{\beta,m}\rangle$, coupled by the Floquet projection operator $\Pi_S$ (i.e., $\langle\psi_{\beta,m}|\Pi_S|\psi_{\alpha,n}\rangle \neq 0$), is larger than $k_r$,

$$|\varepsilon_{\alpha,n} - \varepsilon_{\beta,m}| \gg k_r. \tag{S32}$$

This condition appears to be violated under resonance when $\varepsilon_{\alpha,n} \approx \varepsilon_{\beta,m}$. However, a more detailed analysis shows that level anti-crossing occurs due to the finite dipolar and exchange coupling between spins within the pairs, so that $|\varepsilon_{\alpha,n} - \varepsilon_{\beta,m}| \sim |D|, |J|$. Hence, the condition Eq. S32 is satisfied and the perturbative result Eq. 3 is accurate over the entire parametric domain, provided that $k_r \ll |D|, |J|$.

## S1.5 Numerical evaluation of the spin-dependent current

The starting point of our numerical procedure is a *Monte Carlo* sampling of the set of random quantities entering the stochastic Liouville equation S4, for each individual electron-hole spin pair. The set accounts for local electron and hole hyperfine-field strengths of Gaussian distributions centered around zero, and a random direction for the vector connecting the electron and hole sites, i.e., a unit vector distributed uniformly in all directions. These random quantities form the Floquet Hamiltonian $H_F$ through Eqs. S3, S9, and 1. To this set we add a randomly generated direction for the local axis of quantization, defining the triplet recombination through Eq. S33 of Section S1.6 (another uniformly distributed unit vector).

For the given set of random quantities, we calculate the four matrix elements $\langle\psi_{\alpha,0}|\Pi_S|\psi_{\alpha,0}\rangle$, $\alpha = T_+, T_0, T_-, S$, compute $\text{tr}\tilde{\rho}_0$ from Eq. 3, and average the resulting $\text{tr}\tilde{\rho}_0$ over the random configurations. This calculation is performed after truncating the infinite-dimensional Floquet Hamiltonian $H_F$ and the corresponding Hilbert space. The truncation we employ consists of restricting $n$ and $m$ in Eq. 1 to run from $-N_0$ to $N_0$, where $N_0$ is a suitably chosen integer. Thus, the truncated Hilbert space is $4(2N_0 + 1)$-dimensional, spanned by the dressed states $|\alpha, n\rangle$ with $-N_0 \leq n \leq N_0$. Consequently, the truncated Hamiltonian $H_F'$ and the projection operator $\Pi_S'$ are $4(2N_0 + 1) \times 4(2N_0 + 1)$ matrices.

The eigenvectors $|\psi'_{\alpha,0}\rangle$ found from the numerical solution of the truncated eigenvalue equation $H'_F|\psi'_{\alpha,n}\rangle = \varepsilon'_{\alpha,n}|\psi'_{\alpha,n}\rangle$ are used in Eq. 3 to calculate $\mathrm{tr}\tilde{\rho}_0$. The sampling procedure is repeated $N_s$ times and the observable $I(B_0, B_1)$ sought is found as the average $\langle\mathrm{tr}\tilde{\rho}_0\rangle$. This procedure also controls the choice of $N_0$ and $N_s$ since the truncation size $N_0$ is set by inspecting the convergence with increasing $N_0$ of the truncated matrix element $\langle\psi'_{\alpha,0}|\Pi'_S|\psi'_{\alpha,0}\rangle$ entering into Eq. 3. We verify that, in the parameter space of interest, $N_0 = 4$, corresponding to a $36 \times 36$ Hamiltonian $H'_F$, ensures acceptable convergence. This convergence is also related to the fact that the above matrix element arises between the eigenvectors $|\psi'_{\alpha,0}\rangle$ possessing the largest components $\langle\beta, m|\psi'_{\alpha,0}\rangle$ at $m \sim 0$, while the components at $m \sim \pm N_0$, which are more sensitive to the truncation procedure, have a minimal contribution to $\langle\psi'_{\alpha,0}|\Pi'_S|\psi'_{\alpha,0}\rangle$. The number of sampling steps, $N_s$, is chosen from considerations of accuracy of the *Monte Carlo* averaging. We find that $N_s = 10^5$ provides satisfactory accuracy, within an error of less than 2% of the output.

The specific parameters used in the simulations are listed in Table S1.1. The numerical values of the two combinations of rate constants governing the steady-state solution of the stochastic Liouville equation, $w_d \equiv r_d + r_T + 1/T_{sl}$ and $k_r \equiv r_S - r_T$, are in good agreement with previous experimental results[5]. The rate constant $\delta r_T$ describing the fine structure of the triplet recombination ensures the accurate reproducibility of the $g \approx 4$ resonance feature observed in the experiment (see Section S1.6). Characteristic magnitudes of hyperfine fields $b_{\mathrm{hf},e}$ and $b_{\mathrm{hf},h}$ are equivalent to the standard deviations $\Delta B^D_{\mathrm{hyp},1}$, $\Delta B^D_{\mathrm{hyp},2}$ inferred from the analysis of the power-broadening of experimental line shapes in Section S4. A more accurate, multi-frequency power broadening analysis carried out for the same material[6] yields the parameter values listed in Table S1.1.

| $w_d$(kHz) | $k_r$(kHz) | $\delta r_T$(kHz) | $b_{\mathrm{hf},e}$(mT) | $b_{\mathrm{hf},h}$(mT) | $D$(mT) | $J$(mT) |
|---|---|---|---|---|---|---|
| 290.1 | 58.8 | 21 | 0.076 | 0.244 | 0.22 | 0.065 |

**Table S1.1:** *Parameter values used in the simulations of spin-dependent current.*

**S1.6 Amendments to the conventional spin-pair model**

The above elaborations are based on the conventional spin-pair model of spin-dependent processes in organic semiconductors, see, e.g., Boehme & Lips[2] and Mkhitaryan *et al.*[3]. Small modifications to this model are needed to rationalize the observations made here. The $g \approx 4$ feature seen clearly in our experiment demonstrates a resonant transition between the triplet states $T_+$ and $T_-$. This single-photon transition with the magnetic quantum number changing by $\Delta m = 2$ is possible due to the mixing of the spin-pair states by dipolar interactions[7]. However, under conditions precluding thermal spin polarization and with the assumption of equal dissociation and recombination rates from all triplet states, the states $T_+$ and $T_-$ are populated equally. In this case, none of the spin-pair state populations are affected by a resonant interconversion between $T_+$ and $T_-$ and therefore this resonance cannot induce a change in the device current. This absence of a detectable resonance is confirmed by our simulations within the conventional spin-pair model described above.

The experimentally observed $g \approx 4$ resonance feature is reproduced within the spin-pair model following a modification of the triplet recombination rates. The modification stems from the following arguments. In the course of triplet recombination, a weakly coupled electron-hole spin pair recombines into a triplet exciton. The exciton created is a triplet with zero-field principal axes determined by its local molecular environment. The resulting triplet exciton wavefunctions are quite different for the different spin projection states defined relative to the system of local principal axes. Hence, it is natural to expect that the recombination into these different states can occur with different rates.

We now assume that the rate of triplet spin-pair recombination depends on the specific spin projection state relative to the system of principal axes. Thus, instead of the single recombination rate $r_T$, we employ three different constants $r_{\tilde{T}_+}$, $r_{\tilde{T}_0}$, $r_{\tilde{T}_-}$ for the rates of recombination from the three triplet states in the local principal-axes system $\tilde{T}_+$, $\tilde{T}_0$, and $\tilde{T}_-$. We further find that the experimentally observed resonance features are reproduced quite accurately by setting

$$r_{\tilde{T}_0} = r_T, \qquad r_{\tilde{T}_+} = r_T + \delta r_T, \qquad r_{\tilde{T}_-} = r_T - \delta r_T, \qquad (S33)$$

where $\delta r_T \approx 0.072 w_d$. Remarkably, this modification leaves the *average* recombination rates unchanged, so that the recombination rates from the triplet states $T_+$, $T_0$, and $T_-$ average to the same value, $r_T$. Note that the states $\tilde{T}_+$, $\tilde{T}_0$, and $\tilde{T}_-$ are linear combinations of $T_+$, $T_0$, and $T_-$.

The modified triplet recombination is easily incorporated in Eq. S31 and subsequently in the resulting expression for the spin-dependent current, Eq. 3. We introduce the difference of the projection operators onto the local $\tilde{T}_+$ and $\tilde{T}_-$ states, $\Delta P = |\tilde{T}_+\rangle\langle\tilde{T}_+| - |\tilde{T}_-\rangle\langle\tilde{T}_-|$. In the laboratory basis $T_+, T_0, T_-, S$, this operator has the matrix representation

$$\Delta P = \begin{pmatrix} \cos\theta & \frac{1}{\sqrt{2}}\sin\theta\, e^{i\phi} & 0 & 0 \\ \frac{1}{\sqrt{2}}\sin\theta\, e^{-i\phi} & 0 & \frac{1}{\sqrt{2}}\sin\theta\, e^{i\phi} & 0 \\ 0 & \frac{1}{\sqrt{2}}\sin\theta\, e^{-i\phi} & -\cos\theta & 0 \\ 0 & 0 & 0 & 0 \end{pmatrix}, \tag{S34}$$

where $\theta \in (0, \pi)$ and $\phi \in (0, 2\pi)$ are the spherical angles characterizing the (random) local principal axes system. The Floquet extension of this operator, $\Delta\Pi = \sum_n \left( |\tilde{T}_+, n\rangle\langle\tilde{T}_+, n| - |\tilde{T}_-, n\rangle\langle\tilde{T}_-, n| \right)$, has the matrix representation, $\langle\alpha, n|\Delta\Pi|\beta, m\rangle = \Delta P_{\alpha\beta}\delta_{nm}$. With the rates modified by Eq. S33 we get a modified expression of Eq. 3,

$$\text{tr}\tilde{\rho}_0 \simeq \frac{G}{4}\sum_\alpha \frac{1}{w_d + k_r\langle\psi_{\alpha,0}|\Pi_S|\psi_{\alpha,0}\rangle + \delta r_T\langle\psi_{\alpha,0}|\Delta\Pi|\psi_{\alpha,0}\rangle}. \tag{S35}$$

According to this relation, the set of random quantities from which $\text{tr}\tilde{\rho}_0$ is calculated and subsequently averaged as described in the previous section must be extended to include a pair of spherical angles specifying a system of local principal axes oriented uniformly in all directions. With such an extension and by utilizing the obvious truncation of $\Delta\Pi$, the numerical procedure for the evaluation of the spin-dependent current basically repeats the steps described in the previous section.

**S1.7 Intuitive rationalization of two-photon transitions**

Two-photon transitions between the up and down state of a spin-½ species are dipole forbidden. In analogy to light waves, the resonant radiation can be described in the photon picture. A photon has an angular momentum quantum number of 1 and can assume one of the two projection states of either $m = +1$ or $-1$ relative to the direction of propagation. These are called $\sigma^+$ and $\sigma^-$ photons and can be associates with the right and left circular fields. In a $\sigma$-photon state the AC magnetic field component $B_1$ is perpendicular to the direction of propagation, i.e., $B_1 \perp B_0$, if the direction of propagation is along the quantization axis. The absence of a photon state with $m = 0$ can be traced back to a photon having zero mass. However, a linear AC field $B_1$ in the direction of quantization, $B_1 \parallel B_0$, is referred to as a "$\pi$-photon" propagating perpendicular to the quantization axis. These unconventional $\pi$-photons feature in atomic spectroscopy but are also discussed within a semi-classical picture of magnetic resonance[7,8]. Two-photon transitions are only possible with a combination of $\sigma$ and $\pi$-photons. Since the excitation scheme employed here only involves $\sigma$-type photons, it is not immediately obvious how angular momentum is conserved to give rise to the strong two-photon resonances observed here in theory and experiment. Figure S1.1 provides an intuitive rationalization for the emergence of the two-photon transition. Transversal oscillating magnetic-field components are orthogonal to $B_0$, but the latter is superimposed with local static isotropic hyperfine fields. This superposition leads to the effective static magnetic field, $B_s$, tilted from the $z$-axis by a small but finite angle. Thus, the superposition of $B_0$ and the local hyperfine fields results in an oscillating magnetic-field component parallel to the total static field $B_s$, which is sufficient to enable the $\sigma - \pi$ two-photon transition conserving angular momentum.

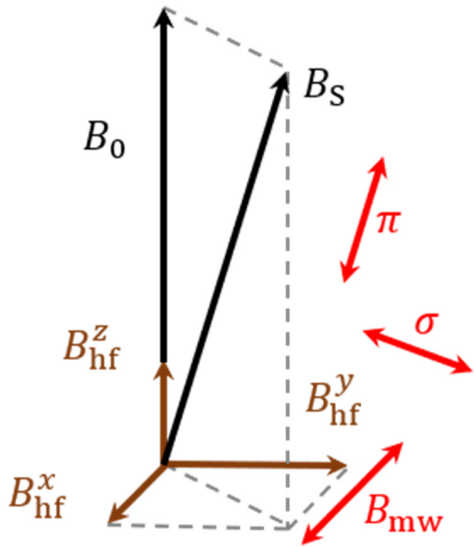

**Figure S1.1.** *Sketch of the magnetic field configuration acting on a spin within a spin pair. Oscillating magnetic field components parallel (π) and perpendicular (σ) to the effective local quantizing static magnetic field $B_S$ arise from a superposition of the external field $B_0$ and the local hyperfine (hf) fields, and the incident microwave (mw) field.*

Furthermore, the magnetic dipolar coupling of spins within the spin pairs also gives rise to a $\pi$-photon component. This becomes clear by considering the dipolar coupling as an effective (nearly static) magnetic field having components perpendicular to $B_0$, acting just like the hyperfine fields in Fig. S1.1. In numerical simulations, it is easy to separate the hyperfine and dipolar contributions in the two-photon resonance. We find that within the parameter domain of interest, see Table S1.1, the two contributions are of similar magnitude.

## S2. Characterization of polymer OLEDs

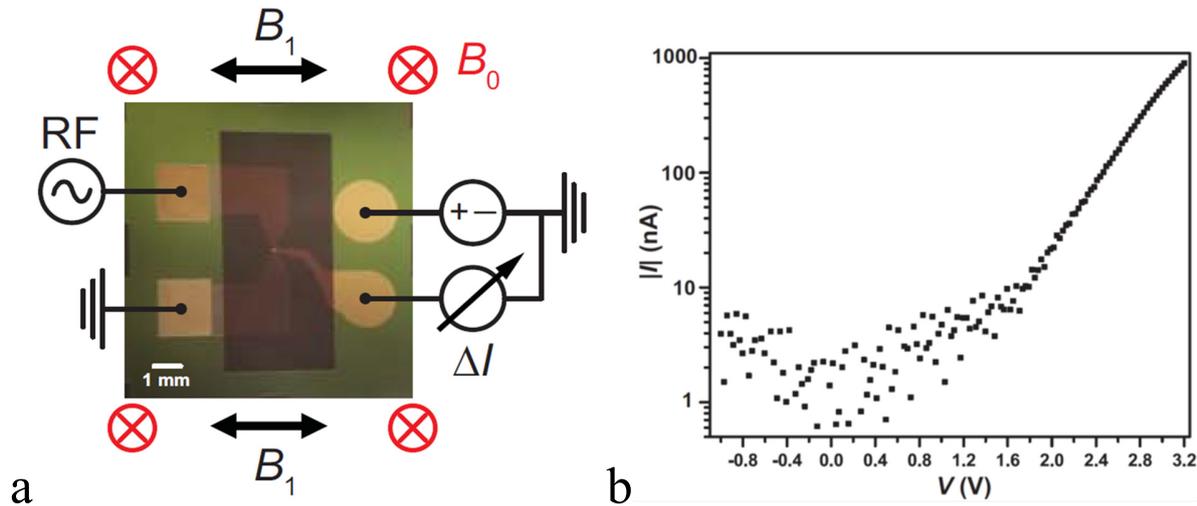

**Figure S2.1. a** *Photograph of a high-$B_1$ monolithic thin-film wire/OLED device with diagrams illustrating the electric circuitry used for the RF excitation as well as the measurement of the device current. The RF field polarization generated by the thin-film wire of amplitude $B_1$ (black arrows) is oriented within the sample plane. The externally applied static magnetic field $B_0$ is oriented perpendicular to the sample plane. The active OLED pixel of diameter 57 μm is seen as the bright spot at the center of the image.* **b** *Current-voltage (I-V) characteristics of the micronscale OLED device.*

## S3. Current change measurement as a function of $B_0$ and RF-power $P \propto B_1^2$

The silicon wafer accommodating the monolithic EDMR-microwire/OLED pixel was placed on a chip holder while electrical connections were established by gold wires and indium dots. The chip holder was placed on top of an aluminum chassis. This holder was then placed in the center of a Helmholtz coil setup for the application of the static $B_0$ field controlled by a Kepco ATE100-10M constant-current source. An RF signal was generated by an Agilent N5181A frequency generator and amplified by an ENI 510L RF power amplifier (9.5 W linear, 45 dBm, 1.7-500 MHz), the output of which was directly connected to the microwire. The largest values of the amplitude $B_1$ of the RF radiation were generated across the narrowest segment of the microwire, right beneath the active device area. The device was biased using a Keithley 2400 source meter and the current was converted by a Stanford Research current preamplifier SRS570 (500 nA offset, 500 nA/V sensitivity with a 10 Hz low-pass filter) into a voltage fed into an NIPCI-6251DAQ analog/digital converter. Data acquisition took place using MATLAB. The steady-state device current was recorded for various values of $B_0$ with and without an 85 MHz RF field applied at different output powers $P$ of the RF generator. Due to the amplifier employed and the non-critical coupling of amplifier and thin-film wire, an arbitrary but constant relationship $B_1 = c\sqrt{P}/2$ between $P$ and $B_1$ exists as long as Ohmic heating effects, which affect the resistance of the Cu microwire, are negligible. The measurements were conducted at room temperature while the sample was operated in an inert atmosphere with a slow flow of $N_2$ gas. This conversion factor $c$ was determined by considering the power broadening of the resonance spectra as described in Section S4. The conversion factor was further corroborated using the Bloch-Siegert shift (BSS) as described in sections S5 through S7.

## S4. Determination of $\sqrt{P}$ to $B_1$ conversion factor through power broadening

The conversion factor between applied power and $B_1$ was determined through measurements of power broadening (PB) of the EDMR resonance line at low driving powers. Figure S4.1 displays a set of EDMR spectra measured for various applied RF powers. The data was fitted globally, i.e., the $B_1$-dependent resonance line shape was fitted to all data sets simultaneously, using three fit parameters: (i) the conversion factor $c_{PB} = \frac{2B_1}{\sqrt{P}}$ and (ii) constant offsets determined by the intrinsic, hyperfine-field governed linewidths of the two carrier-pair resonances which define the resonance line width in the absence of power broadening. The precise numerical procedure used is described in detail elsewhere[9]. The results of this procedure are $c_{PB} = 0.0217(60)$ mT/$\sqrt{mW}$ as well as $\Delta B_{hyp,1}^{D} = 0.081(2)$ mT for the narrow hole charge-carrier resonance and $\Delta B_{hyp,2}^{D} = 0.28(20)$ mT for the broad electron charge-carrier resonance line. Note that these values represent the standard deviation of the Gaussian spin-resonance spectra as defined in Joshi *et al.*[10] and Malissa *et al.*[11]; they do not represent the full width at half maximum (FWHM) of the charge-carrier spin-resonance spectral lines as discussed by Waters *et al.*[12], which are larger than the standard deviations by a factor of $2\sqrt{2\ln2}$.

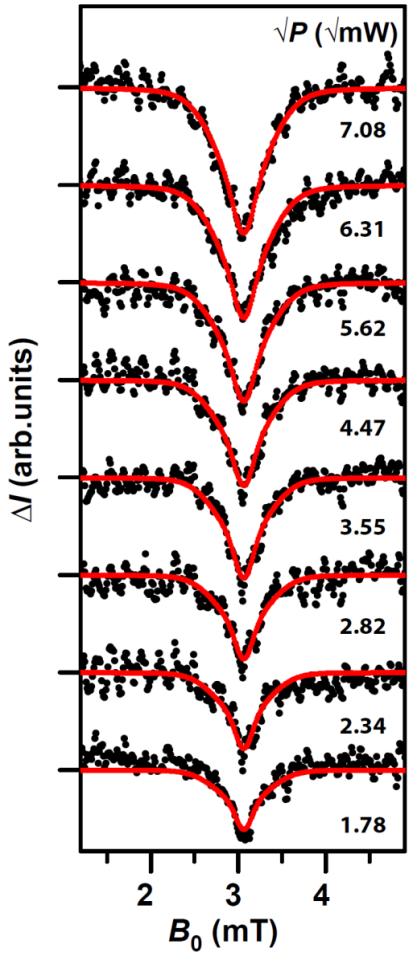

**Figure S4.1.** *Black dots: plots of OLED current change ΔI as a function of the applied magnetic field $B_0$ in the presence of an 85 MHz RF radiation field at various applied RF powers as indicated for each data set. Red lines: simultaneous global fits of all experimental data sets using a model in which the resonance line width is governed by power broadening and the intrinsic Gaussian hyperfine distributions experienced by electron and hole spins.*

## S5. Hyperfine-field controlled line widths of the spin-pair resonance

The widths of the two low-frequency (<1 GHz) low-power EDMR resonances in MEH-PPV are known to be determined by the expectation values of the two local hyperfine-field distributions $\Delta B_{\text{hyp},1}^{\text{D}}$ and $\Delta B_{\text{hyp},2}^{\text{D}}$ [6,9,10-12]. In fact, the widths of the two carrier resonance lines reported here compare to those reported previously for fully protonated h-MEH-PPV, $\Delta B_{\text{hyp},1}^{\text{H}} = 0.2080(8)$ mT for the narrow hole resonance and $\Delta B_{\text{hyp},2}^{\text{H}} = 0.8111(27)$ mT for the broad electron resonance [10,11], according to the ratios of $\frac{\Delta B_{\text{hyp},1}^{\text{D}}}{\Delta B_{\text{hyp},1}^{\text{H}}} = 0.389(9)$ and $\frac{\Delta B_{\text{hyp},2}^{\text{D}}}{\Delta B_{\text{hyp},2}^{\text{H}}} = 0.34(25)$. While the latter shows reasonable agreement with the ratio $\mu_{\text{D}}/\mu_{\text{H}} = 0.3070121$ of the nuclear magnetons of deuterium and protium, the former is slightly higher, likely because other inhomogeneous line-broadening effects may have influenced the narrow resonance line of the deuterated material under these conditions of very low hyperfine fields.

## S6. Determination of $\sqrt{P}$ to $B_1$ conversion factor by the Bloch-Siegert shift

The detection of the spin-Dicke effect can be used in order to observe another signature of ultra-strong coupling, namely the breakdown of the rotating-wave approximation for the description of light-matter interaction. The physical origin of the latter, which expresses itself by a shift of the resonance frequency with increasing drive strength, the Bloch-Siegert shift (BSS)[13], is illustrated in Fig. S6.1. The graph shows plots of the transition probability

$$w_{\downarrow\to\uparrow}(B_0) = \overline{|\langle\psi(t)|\uparrow\rangle|^2} = \frac{\gamma B_1}{2\Omega_\pm(B_0)}, \qquad (S36)$$

representing the time average of Rabi's formula[13,14] from the ground state $|\downarrow\rangle$ of a $s = \frac{1}{2}$ electron spin to the excited state $|\uparrow\rangle$ as a function of an applied static magnetic field $B_0$ for the conditions of ultrastrong and weak resonant drive. The red line indicates right-handed helicity of the RF radiation, generating a Rabi frequency $\Omega_+$, with blue labelling the left-handed helicity generating a Rabi frequency $\Omega_-$. $\gamma = 28.03$ GHz/T is the electron spin's gyromagnetic ratio, which is related to the g-factor through $\gamma = g\mu_B/\hbar$, and

$$\Omega_\pm(B_0) = \sqrt{(\gamma B_1)^2 + (\gamma B_0 \mp \omega)^2} \qquad (S37)$$

are the Rabi frequencies of the two helicity components with $\omega = 2\pi f$ being the frequency of the applied radiation, assuming that there is no spin-orbit coupling, local random hyperfine interactions, coupling to other spin systems, or coupling that could cause spin relaxation. The plots in Fig. S6.1 are based on RF driving fields with $f = 85$ MHz, and $B_1 = 3.5$ mT and 0.5 mT radiation for the strong and weak-drive cases, respectively. $B_1$ is linearly polarized in an $\hat{x}$-$\hat{y}$ plane, while $B_0$ is parallel to $\hat{z}$. Figure S6.1 also displays normalized sums of the blue and red plots (black lines), i.e., superpositions of the two helicities, describing the case of linearly polarized RF radiation.

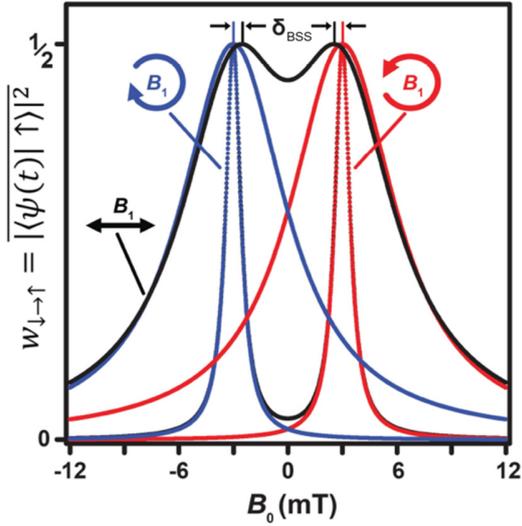

**Figure S6.1.** *Illustration of the spin-dependent transition-rate changes induced by magnetic resonance for weak (narrow peaks) and strong (broad peaks) electromagnetic drive conditions under linearly polarized excitation (black lines) and under circularly polarized excitation (red, blue). The simulated line shapes are normalized.*

When the inhomogeneous broadening due to, e.g., the hyperfine-field distribution, is small compared to $B_1$, the line shape described by Eq. S34 will be predominantly Lorentzian with a full-width-at-half-maximum (FWHM) of $2B_1$, an effect referred to as power broadening. When $B_1 \ll B_0$, distinct narrow resonance lines emerge whose centers and shapes are nearly equal for linear and circular polarization states, even though for the circularly polarized excitation, the resonance peaks occur only when the field direction is positive for the right-handed helicity and negative for the left-handed helicity, i.e., for the EPR-active helicities as shown in Fig. S6.1. This behavior is explained by the rotating-wave picture where the Rabi precession of the electron spin nutates due to a constant magnetic field $B_1$ along the $\hat{x}$-axis in a reference frame which rotates along with the EPR-active helicity of the circularly polarized light when the spin's Larmor frequency $\omega_0 = \gamma B_0$ and the radiation frequency $\omega$ are identical, i.e., in resonance. No such nutation can take place for the EPR-inactive helicity since $\omega_0 \neq -\omega$. The rotating-frame picture is not only illustrative, its mathematical implementation in the form of the rotating-wave approximation also allows the time-dependent perturbation by circularly polarized light to render the spin-Hamiltonian an exactly solvable problem. Under application of linear polarization, resonance peaks occur in both positive and negative field directions as the linear polarization state is the superposition of the two circular polarization states. Under weak drive, the resonance peaks for both positive and negative static field values shown in Fig. S6.1 are identical, and thus linearly polarized excitation can be approximated by a rotating wave. This approximation is equivalent to circularly polarized excitation with positive helicity under complete neglect of the out-of-resonance helicity.

Under ultrastrong drive, power broadening becomes large enough such that the contributions of the EPR-active and EPR-inactive helicities superimpose. The rotating-wave approximation then breaks down, causing a BSS of the local frequency maximum (the peak center frequency) $\delta\omega_0 = \frac{(\gamma B_1)^2}{4\omega_0}$ relative to the Larmor frequency[15,16]. The positive BSS of the resonance frequency with increasing $B_1$ causes a negative BSS of the center magnetic field $B_c$

$$\delta_{BSS} = \frac{B_c}{2} - \sqrt{\left(\frac{B_c}{2}\right)^2 - \left(\frac{B_1}{2}\right)^2} \qquad (S38)$$

under experimental conditions where a constant drive-field frequency $f$ is applied and the magnetic field is swept as seen in the illustration in Fig. S6.1 for the broad resonance lines. Detection of the BSS through the shift of the center of the magnetic resonance line is therefore possible, but only within the ultrastrong drive regime (as the BSS is too small when $B_1 \ll B_0$). Furthermore, in the regime of the so-called deep-strong drive where $B_1 > B_0$, the resonance maximum will become pinned at $B_0 = 0$. Within the intermediate regime of strong drive, the BSS becomes progressively more pronounced, while the resonance peak becomes smaller due to the effect of power broadening (cf. Fig. S4.1). An accurate determination of the peak center as needed to quantify the BSS then becomes difficult with conventional magnetic resonance spectroscopy.

It is the particular signature of spin-collectivity in the device current on resonance, the sign reversal of spin-dependent current changes under strong drive conditions, which allows us to circumvent this problem arising from power broadening. Since the spin-Dicke effect inverts $\Delta I$ on resonance only, i.e., within a narrow range $|\omega - \gamma B_0| \lesssim \mu_B B_{\text{hyp}}/\hbar$ around the resonance condition, it is not subject to power broadening. Thus, while power broadening increases with $B_1$, the spin-Dicke effect remains an indicator for the resonance peak center with an accuracy that is effectively independent of $B_1$. Indeed, the data in Fig. 3 of the main text shows that the width of the inverted current change $\Delta I(B_0)$ due to the spin-Dicke effect [feature (i) in Fig. 3 of the main text] remains almost unchanged as $B_1$ is increased, in contrast to the non-inverted conventional carrier-pair spin resonance, which shows strong power broadening. The current changes measured at the highest driving powers corresponding to $B_1 > 2$ mT show such strong power broadening—the blue regions in Fig. 3 represent current quenching due to very broad EDMR peaks—that the peak center cannot be determined with sufficient accuracy to corroborate the BSS. Only with the spin-Dicke effect as observed in Fig. 3 [feature (i)] a clear shift of the resonance center of the red area in the vertical ($B_0$) direction towards lower values of $B_0$ can be identified.

Figure S6.2a displays several selected experimental magnetoresistance data sets $\Delta I(B_0)$ extracted from the data of Fig. 3 of the main text for various values of $P$. In order to exploit the quantitative nature of the BSS for the determination of the $\sqrt{P}$ to $B_1$ conversion, all extrema for all spectra were determined as a function of $\sqrt{P}$ using a standardized procedure. To find the center values of magnetic field $B_0^c$ of the $g \approx 2$ resonance for each applied RF power $P$, the data for each spectrum $\Delta I(B_0)$ was subjected to a 40 data-point smoothing procedure using the *Origin* Software *FFT smoothing method* (https://www.originlab.com/doc/Origin-Help/Smoothing). Subsequently, the package's peak analyzer was used to find the absolute maximum defining $B_0^c$ for each value of $P$. The resulting dependency of $B_0^c(\sqrt{P})$ shown in Fig. S6.2b was then fitted with the expression

$$B_0^c(\sqrt{P}) = \frac{B_0^c(0)}{2} \pm \sqrt{\left[\left(\frac{B_0^c(0)}{2}\right)^2 - \left(\frac{B_1}{2}\right)^2\right]} \qquad (S39)$$

in which $B_1 = \frac{c_{BSS}\sqrt{P}}{2}$ and $B_0^c(0) = f/\gamma = 3.0325$ mT was defined by the gyromagnetic ratio $\gamma = 28.03 \frac{GHz}{T}$ and the applied radiation frequency $f = 85$ MHz. The fit was therefore reduced to a single fit variable, namely the conversion factor $c_{BSS} = \frac{2B_1}{\sqrt{P}} = 0.018(2)$ mT/$\sqrt{mW}$.

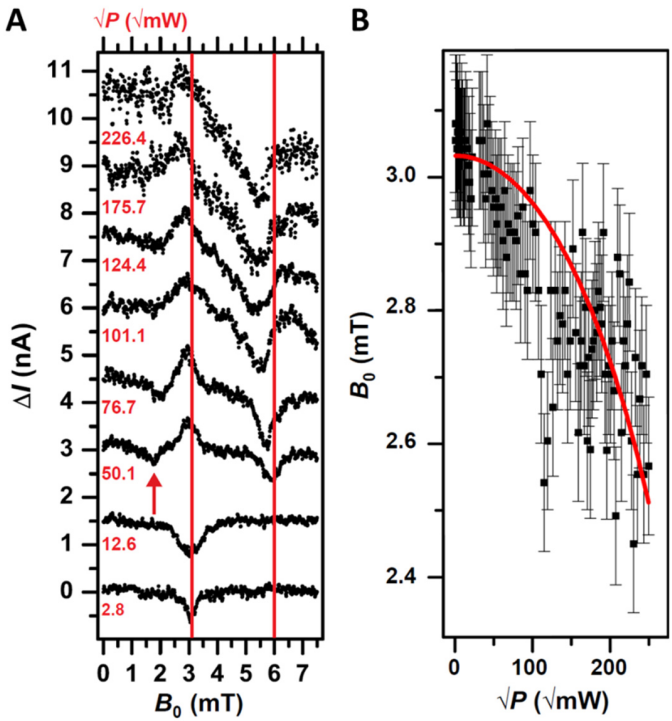

**Figure S6.2. a** *Plot of the current change ΔI as a function of the static magnetic field $B_0$ in the presence of RF radiation with different applied powers as indicated for each data set (red labels), corresponding to different driving-field amplitudes $B_1$. The red lines indicate the resonance centers of the $g \approx 2$ and the $g \approx 1$ lines for small $B_1$. The arrow indicates the half-field resonance.* **b** *Black points: plot of the resonance-peak extrema of the data shown in panel **a** as a function of the applied driving field amplitude $B_1 = c\sqrt{P}$. Red line: fit of the data using a model function based on the BSS.*

The standard deviation (the error) for $B_0^c$ was estimated on the basis of the residuals of the set of peak values $B_0^c(\sqrt{P})$ obtained by the procedure described above and the fit results based on Eq. S37. The distribution of these residuals is shown in the histogram in Fig. S6.3. Using this distribution and an estimator for an unbiased sample variance, a standard deviation of 0.106 mT was found for $B_0^c$. We note that the center of this distribution has a slight offset of -0.0417 mT, which is due to an emphasis of the residuals of the fit function for powers in the range of $40 \sqrt{mW} \leq \sqrt{P} \leq 100 \sqrt{mW}$. This offset is caused by the fact that the fit model does not account for the half-field resonance signal (red arrow in Fig. S6.2), affecting the precision in determining $B_0^c$.

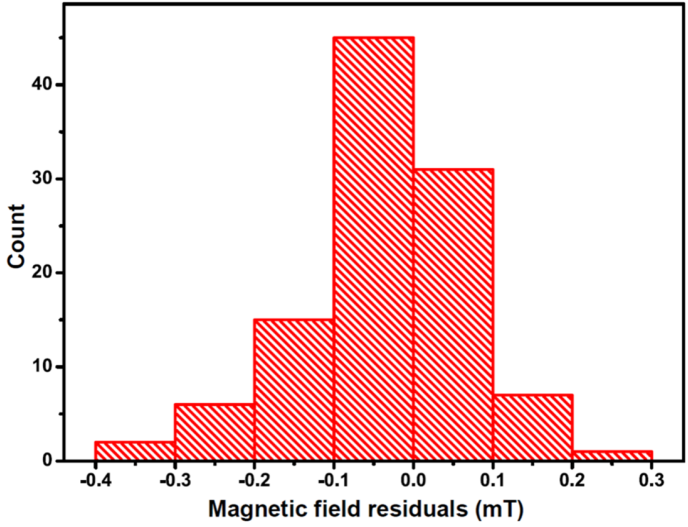

***Figure S6.3.*** *Histogram showing the number of counts for the fit residuals of the magnetic-field peak center value $B_0^c$ from fitting to Eq. S37.*

## S7. Spin-Dicke effect of perdeuterated d-MEH-PPV

Figure S7.1 displays the plot of the on-resonance ($g \approx 2$) current change normalized to its minimum as a function of the driving-field amplitude $B_1$, and scaled along the abscissa to the local hyperfine-field strength $B_{hyp}$ experienced by the charge carrier for fully protonated (black), perdeuterated (red), and partially deuterated[12], (blue) MEH-PPV. The increase of this function after an initial linear decrease is characteristic for the onset of the spin-Dicke effect[17], and these functions are predicted to solely scale with $B_{hyp}$. For the plots shown in Fig. S7.1, the scaling factors of $B_{hyp}^H = 0.97$ mT and $B_{hyp}^{DH} = 0.51$ mT for the fully protonated and mixed MEH-PPV were taken from Waters *et al.*[12], while the scaling factor $B_{hyp}^D = 0.24(3)$ mT for perdeuterated MEH-PPV was obtained from the data presented here as discussed above in Section S5. This value was obtained through adjustment of the scaling factor such that maximal overlap with the other two functions occurred under the assumption that the scale for $B_1$ is given by the power conversion factor $c_{BSS}$ as obtained from the fit of the BSS discussed in Section S6. This procedure leads to a ratio $B_{\text{hyp}}^{\text{D}}/B_{\text{hyp}}^{\text{H}} = 0.25(3)$ of the scaling factors, which is slightly smaller than the ratio $\mu_{\text{D}}/\mu_{\text{H}} = 0.3070121$ of the nuclear magnetons of deuterium and protium.

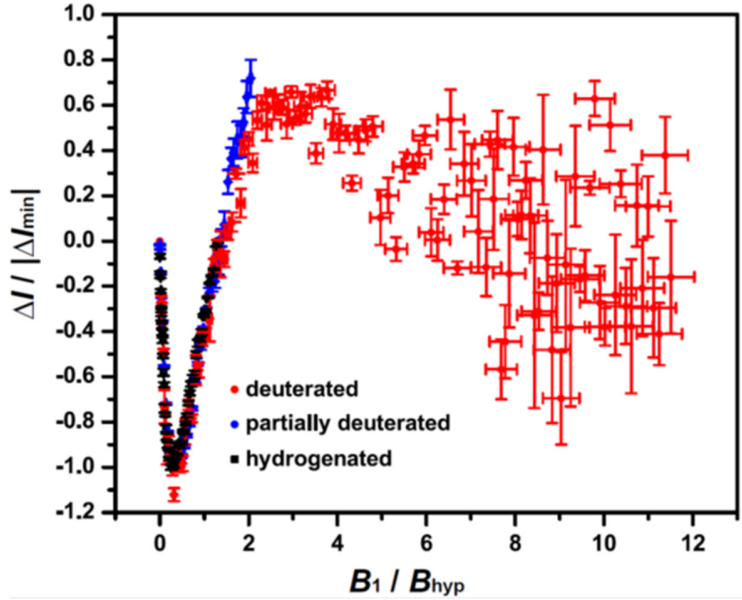

**Figure S7.1.** *Plot of the on-resonance current change ($B_0 = 3.05$ mT, $f = 85$ MHz) normalized to the current change when the current minimum occurs, as a function of the amplitude of the driving field $B_1$ normalized to the expectation value of hyperfine field strength for three different OLED samples. The three data sets correspond to devices with hydrogenated, partially deuterated (both taken from Waters* et al.[12]*) and fully deuterated MEH-PPV.*

The experiments on perdeuterated MEH-PPV reported here provide significantly improved ratios of $B_1/B_{\text{hyp}}$ in comparison to earlier work[9,12]. The observation made under these conditions confirms the theoretically predicted saturation behavior of the current change in the spin-Dicke regime beyond the onset of the spin-Dicke effect, i.e., where the sign change of $\Delta I$ occurs[17]. The gradual decrease of the change in saturation current with increasing $B_1$ is a consequence of the BSS, which causes the resonance to detune from the $g \approx 2$ center magnetic field at high driving powers.

**References**

1. van Schooten, K. J., Baird, D. L., Limes, M. E., Lupton, J. M. & Boehme, C. Probing long-range carrier-pair spin-spin interactions in a conjugated polymer by detuning of electrically detected spin beating. *Nat. Commun.* **6**, 6688 (2015).

2. Boehme, C. & Lips, K. Theory of time-domain measurements of spin-dependent recombination with pulsed electrically detected magnetic resonance. *Phys. Rev. B* **68**, 245101 (2003).

3. Mkhitaryan, V. V., Danilović, D., Hippola, C., Raikh, M. E. & Shinar, J. Comparative analysis of magnetic resonance in the polaron pair recombination and the triplet exciton-polaron quenching models. *Phys. Rev. B* **97**, 035402 (2018).

4. Shirley, J. H. Solution of the Schrödinger Equation with a Hamiltonian Periodic in Time. *Phys. Rev.* **138**, B979-B987 (1965).

5. Baker, W. J., Keevers, T. L., Lupton, J. M., McCamey, D. R. & Boehme, C. Slow Hopping and Spin Dephasing of Coulombically Bound Polaron Pairs in an Organic Semiconductor at Room Temperature. *Phys. Rev. Lett.* **108**, 267601 (2012).

6. Stoltzfus, D. M. *et al.* Perdeuteration of poly[2-methoxy-5-(2′-ethylhexyloxy)-1,4-phenylenevinylene] (d-MEH-PPV): control of microscopic charge-carrier spin–spin coupling and of magnetic-field effects in optoelectronic devices. *J. Mater. Chem. C* **8**, 2764-2771 (2020).

7. Schweiger, A. & Jeschke, G. *Principles of Pulse Electron Paramagnetic Resonance*, Oxford University Press (2001).

8. Gromov, I. & Schweiger, A. Multiphoton Resonances in Pulse EPR. *J. Magn. Reson.* **146**, 110-121 (2000).

9. Jamali, S., Joshi, G., Malissa, H., Lupton, J. M. & Boehme, C. Monolithic OLED-Microwire Devices for Ultrastrong Magnetic Resonant Excitation. *Nano Lett.* **17**, 4648-4653 (2017).

10. Joshi, G. *et al.* Separating hyperfine from spin-orbit interactions in organic semiconductors by multi-ocatve magnetic resonance using coplanar waveguide microresonators. *Appl. Phys. Lett.* **109**, 103303 (2016).

11. Malissa, H. *et al*. Revealing weak spin-orbit coupling effects on charge carriers in a π-conjugated polymer. *Phys. Rev. B* **97**, 161201(R) (2018).

12. Waters, D. P. *et al*. The spin-Dicke effect in OLED magnetoresistance. *Nat. Phys.* **11**, 910-914 (2015).

13. Bloch, F. & Siegert, A. Magnetic Resonance for Nonrotating Fields. *Phys. Rev.* **57**, 522-527 (1940).

14. Gross, M. & Haroche, S. Superradiance: An essay on the theory of collective spontaneous emission. *Phys. Rep.* **93**, 301-396 (1982).

15. Malissa, H. *et al*. Room-temperature coupling between electrical current and nuclear spins in OLEDs. *Science* **345**, 1487-1490 (2014).

16. Wei, C., Windsor, A. S. M. & Manson, N. B. A strongly driven two-level atom revisited: Bloch-Siegert shift versus dynamic Stark splitting. *J. Phys. B: At. Mol. Opt. Phys.* **30**, 4877-4888 (1997).

17. Roundy, R. C. & Raikh, M. E. Organic magnetoresistance under resonant ac drive. *Phys. Rev. B* **88**, 125206 (2013).